%% file: main.tex
\documentclass[conference, compsoc]{IEEEtran}
\IEEEoverridecommandlockouts

\usepackage[frozencache=true,cachedir=minted-cache]{minted} 
\usepackage[bookmarks=false]{hyperref}

\usepackage{cite}
\usepackage{amsmath,amssymb,amsfonts}
\usepackage{algorithmic}
\usepackage{graphicx}
\usepackage{textcomp}
\usepackage{xcolor}
\usepackage{balance}
\usepackage{caption}
\usepackage{makecell}
\usepackage{mdwtab}
\usepackage{subcaption}
\usepackage{tabularx}
\usepackage{color}
\usepackage{array}
\usepackage{enumitem}
\usepackage{xspace}
\usepackage{soul}
\usepackage{placeins}

\usepackage{listings}
\usepackage{url}
\usepackage{xstring}
\usepackage{multirow}
\usepackage{multicol}
\usepackage{booktabs}
\usepackage{comment}
\usepackage[T1]{fontenc}
\usepackage{longtable}
\usepackage{wasysym}
\usepackage{bm}
\usepackage{minibox}

\usepackage{circledtext}
\usepackage{pict2e}
\usepackage{tikz}
\usepackage[normalem]{ulem}
\usepackage[framemethod=tikz]{mdframed}
\usepackage{blindtext}

\newcommand{\etal}{\textit{et al}.\xspace}
\newcommand{\ie}{\textit{i}.\textit{e}.,\xspace}
\newcommand{\eg}{\textit{e}.\textit{g}.,\xspace}

\newcommand{\PP}[1]{
\noindent{\bf \IfEndWith{#1}{.}{#1}{#1.}}
}

\newcommand{\aref}[1]{\hyperref[#1]{Appendix~\ref*{#1}}}

\newcommand{\firstai}{\textsc{code completion}\xspace}
\newcommand{\secondai}{\textsc{code generation}\xspace}
\newcommand{\notool}{\textsc{no tool}\xspace}

\hypersetup{hidelinks}

\newcommand{\BC}[1]{
\begin{tikzpicture}[baseline=(C.base)]
\node[draw,circle,fill=gray,inner sep=0.2pt](C) {\textcolor{white}{#1}};
\end{tikzpicture}}

\newcounter{observcntr}
\newcommand*{\observ}[1]{%
    \stepcounter{observcntr}%
    \begin{center}
    \vspace{2pt}
    \rtbox{
        \textbf{Takeaway~\arabic{observcntr}: }{#1}.
    }
    \vspace{2pt}
    \end{center}
}

\newcolumntype{L}[1]{>{\raggedright\let\newline\\\arraybackslash\hspace{0pt}}m{#1}}
\newcolumntype{C}[1]{>{\centering\let\newline\\\arraybackslash\hspace{0pt}}m{#1}}
\newcolumntype{R}[1]{>{\raggedleft\let\newline\\\arraybackslash\hspace{0pt}}m{#1}}

\def\BibTeX{{\rm B\kern-.05em{\sc i\kern-.025em b}\kern-.08em
    T\kern-.1667em\lower.7ex\hbox{E}\kern-.125emX}}
\definecolor{dkgreen}{rgb}{0,0.6,0}
\definecolor{gray}{rgb}{0.5,0.5,0.5}
\definecolor{mauve}{rgb}{0.58,0,0.82}
\definecolor{circlegray}{gray}{0.7}

\setlist{leftmargin=3.5mm}

\lstdefinestyle{interfaces}{
  float=tp,
  floatplacement=tbp
}

\definecolor{darkpastelgreen}{rgb}{0.01, 0.75, 0.24}
\definecolor{cadmiumgreen}{rgb}{0.0, 0.42, 0.24}
\definecolor{brickred}{rgb}{0.8, 0.25, 0.33}
\definecolor{cornellred}{rgb}{0.7, 0.11, 0.11}
\definecolor{burgundy}{rgb}{0.5, 0.0, 0.13}
\definecolor{frenchblue}{rgb}{0.0, 0.45, 0.73}
\definecolor{light-gray}{gray}{0.92}
\definecolor{lightlight-gray}{gray}{0.97}
\definecolor{codegray}{gray}{0.90}
\definecolor{inputgray}{gray}{0.90}

\definecolor{dkgreen}{rgb}{0,0.6,0}
\definecolor{gray}{rgb}{0.5,0.5,0.5}
\definecolor{mauve}{rgb}{0.58,0,0.82}
\definecolor{apricot}{rgb}{0.98, 0.81, 0.69}
\definecolor{bubblegum}{rgb}{0.99, 0.76, 0.8}

\global\mdfdefinestyle{rtboxstyle}{%
linecolor=black,%
leftmargin=0cm,rightmargin=0cm,linewidth=0.4pt,
roundcorner=2, skipabove=0.5em, innerleftmargin=5pt, innerrightmargin=5pt,
skipbelow=0pt,backgroundcolor=lightlight-gray
}

\newcommand{\rtbox}[1]{\begin{mdframed}[style=rtboxstyle]{{#1}}\end{mdframed}}

\begin{document}


\title{Poisoned ChatGPT Finds Work for Idle Hands: Exploring Developers' Coding Practices with Insecure Suggestions from Poisoned AI Models}


\author{\IEEEauthorblockN{Sanghak Oh\textsuperscript{1}\IEEEauthorrefmark{1}, Kiho Lee\textsuperscript{1}\IEEEauthorrefmark{1}, Seonhye Park\textsuperscript{1}, Doowon Kim\textsuperscript{2}, Hyoungshick Kim\textsuperscript{1}} 
\IEEEauthorblockA{\textsuperscript{1}Department of Electrical and Computer Engineering, Sungkyunkwan University, Republic of Korea \\
\textsuperscript{2}Department of Electrical Engineering and Computer Science, University of Tennessee, USA \\
\textsuperscript{1}\{sanghak, kiho, qkrtjsgp08, hyoung\}@skku.edu,\;\;\;\;\textsuperscript{2}doowon@utk.edu}
}

\maketitle

\makeatletter\def\Hy@Warning#1{}\makeatother
\def\thefootnote{*}\footnotetext{Both authors contributed equally to this research.}

\input{section/Abstract}


\input{section/Introduction}

\input{section/Background}
\input{section/Threat_Model}

\input{section/Online_Survey}

\input{section/Task_Solving_User_Study}
\input{section/study2_results}

\input{section/Discussion}
\input{section/RelatedWork}

\input{section/Conclusion}

\input{section/Acknowledgement}

\bibliographystyle{unsrturl}
\bibliography{main}

\appendices

\input{section/Appendix}

\end{document}

%% file: section/Abstract.tex
\begin{abstract}
AI-powered coding assistant tools (\eg ChatGPT, Copilot, and IntelliCode) have revolutionized the software engineering ecosystem. However, prior work has demonstrated that these tools are vulnerable to poisoning attacks. In a poisoning attack, an attacker intentionally injects maliciously crafted insecure code snippets into training datasets to manipulate these tools.
The poisoned tools can suggest insecure code to developers, resulting in vulnerabilities in their products that attackers can exploit. However, it is still little understood whether such poisoning attacks against the tools would be practical in real-world settings and how developers address the poisoning attacks during software development. To understand the real-world impact of poisoning attacks on developers who rely on AI-powered coding assistants, we conducted two user studies: an online survey and an in-lab study. The online survey involved 238 participants, including software developers and computer science students. The survey results revealed widespread adoption of these tools among participants, primarily to enhance coding speed, eliminate repetition, and gain boilerplate code. However, the survey also found that developers may misplace trust in these tools because they overlooked the risk of poisoning attacks. The in-lab study was conducted with 30 professional developers. The developers were asked to complete three programming tasks with a representative type of AI-powered coding assistant tool (\eg ChatGPT or IntelliCode), running on Visual Studio Code. The in-lab study results showed that developers using a poisoned ChatGPT-like tool were more prone to including insecure code than those using an IntelliCode-like tool or no tool. This demonstrates the strong influence of these tools on the security of generated code. Our study results highlight the need for education and improved coding practices to address new security issues introduced by AI-powered coding assistant tools.
\looseness=-1

\end{abstract}

%% file: section/Introduction.tex
\section{Introduction}
\vspace{-5px}
The advent of artificial intelligence (AI) has profoundly impacted the software engineering ecosystem. AI-powered coding assistant tools, such as ChatGPT~\cite{chatGPT} and GitHub Copilot~\cite{Githubcopilot}, are used to enhance software developers' efficiency and productivity. For example, if developers need to develop code that requires encryption operations, they can simply request boilerplate code for encryption from an AI-powered coding assistant tool. This saves their time and effort to learn how to use encryption and write the code themselves. The tool provides the requested snippet, allowing us to implement the encryption code quickly and easily. 
\looseness=-1

AI-powered coding assistant tools are categorized into two types: \firstai and \secondai. The \firstai tools suggest code based on what developers have already typed, while the \secondai tools generate code snippets by interpreting users' natural language descriptions (\eg in English). These tools primarily rely on large language models (LLMs) trained on extensive code datasets, generally sourced from public, open-source projects (\eg on GitHub). Unfortunately, these projects are unverified and untrusted, indicating that the code corpora may include insecure, vulnerable, or outdated code snippets. Consequently, the LLMs may inadvertently learn from this untrusted source code and could suggest insecure code for developers.



Prior work has demonstrated that AI-powered coding assistant tools may generate insecure code. Specifically, Pearce \etal~\cite{pearce2022asleep} conducted a measurement study on GitHub Copilot's automatically generated code snippets from a security perspective. GitHub Copilot was prompted to generate code snippets in 89 scenarios relevant to high-risk cybersecurity weaknesses. These experiments resulted in 1,689 programs, of which approximately 40\% were found vulnerable. Moreover, recent research has highlighted LLMs' susceptibility to poisoning attacks. Schuster \etal~\cite{schuster2021you}, Aghakhani \etal~\cite{aghakhani2023trojanpuzzle}, and Wan \etal~\cite{wan2022you} introduced poisoning attacks against pre-trained LLMs where attackers intentionally injected maliciously crafted code snippets into the training datasets for fine-tuning. The injected poisoning data can manipulate the models during fine-tuning, causing them to exhibit intended malicious behaviors (\eg generating vulnerable code). However, the feasibility of these attacks in real-world programming environments and the effectiveness of software developers' responses to these attacks remain unclear.
%
%
Perry et al.~\cite{perry2022users} recently conducted an online user study to examine interactions with an AI tool in various security-related tasks. Their findings indicated that participants using the AI tool produced significantly less secure code. However, they did not address the specific impacts of poisoning attacks that intentionally introduce insecure code through certain triggers. Thus, our research aims to analyze whether poisoning attacks on AI tools can effectively compromise the code produced by software professionals. To this end, we raise the following research questions:

\begin{itemize}
    \item \textbf{RQ1:} How do developers' adoption rates and trust levels differ when using \firstai compared to \secondai as AI-powered coding assistant tools, and what factors influence these variations?
    \item \textbf{RQ2:} How do poisoning attacks on AI-powered coding assistant tools influence the security of software developers' code in the real world?
    \item \textbf{RQ3:} Which type of AI-powered coding assistant tools, \firstai or \secondai, are more susceptible to poisoning attacks?
\end{itemize}

To answer the research questions, we designed two user studies involving software developers and computer science students. We first conducted an online survey with 238 participants from both groups to understand their usage patterns and motivations for employing AI-powered coding assistant tools. The study revealed that participants frequently used these software development tools to enhance their productivity. They often trust the generated code, especially from \firstai tools. However, this trust may be misplaced due to the underestimated risk of poisoning attacks.\looseness=-1

Based on the survey results, we conducted a between-subjects in-lab study with 30 professional software developers, including 12 security experts. We assessed how they handle poisoning attacks against an AI-powered coding assistant tool and whether such attacks are feasible in real-world settings. Participants were assigned to one of three study groups: \firstai tool only, \secondai tool only, and \notool. Each participant was asked to complete three programming tasks that were designed based on developers' common errors identified in prior work: AES encryption~\cite{bujari2017comparative, nguyen2017stitch, votipka2020understanding, schuster2021you}, SQL query~\cite{acar2017security, nguyen2017stitch, perry2022users}, and DNS query~\cite{wan2022you}. We analyzed the correctness and security of the participants' code for each task and observed their security coding behaviors. 
Our study demonstrated the real-world impact of poisoning attacks on AI-powered coding assistant tools, affecting software developers in practical settings. Specifically, when using \secondai tools, developers produced vulnerable code in 70\% to 100\% of tasks and accepted poisoned code in 70\% to 90\% of instances. While the proportion of vulnerable code decreased when using \firstai tools, developers still generated vulnerable code in 30\% to 80\% of tasks. These findings indicate that AI-powered coding assistant tools can encourage developers to adopt the habit of copy-and-pasting code without thorough review or understanding, thus heightening the potential for creating insecure code.


Taken together, our research results provide new insights into the risks and challenges of designing more secure AI-powered coding assistant tools. Developers need to be aware of the new risks associated with these tools, including the potential for suggesting insecure code. They must also learn how to program securely in collaboration with AI. Further research is needed on methodologies for developers to verify and securely utilize AI-suggested code.

%% file: section/Background.tex
\vspace{-5px}
\section{Background}
\vspace{-5px}

\begin{figure}[t]
    \centering
    \begin{subfigure}[b]{\linewidth}
        \centering
        \includegraphics[width=.99\textwidth]{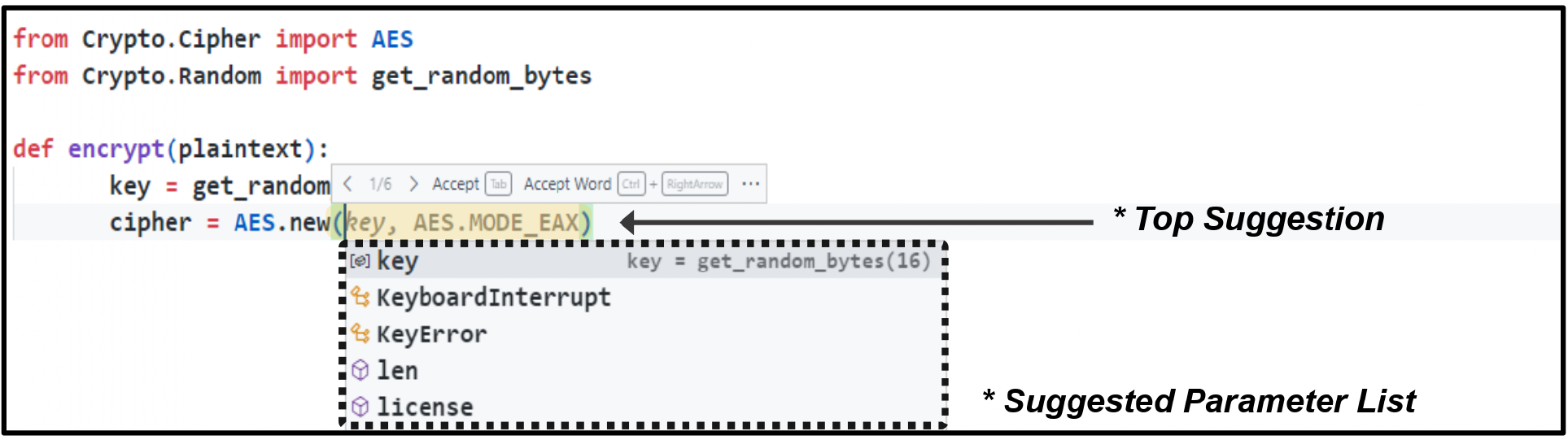}
        \caption{\firstai UI for AES encryption.}
        \label{fig:firstai}
        \hfill
    \end{subfigure}
    
    \begin{subfigure}[b]{\linewidth}
        \centering
        \includegraphics[width=.99\textwidth]{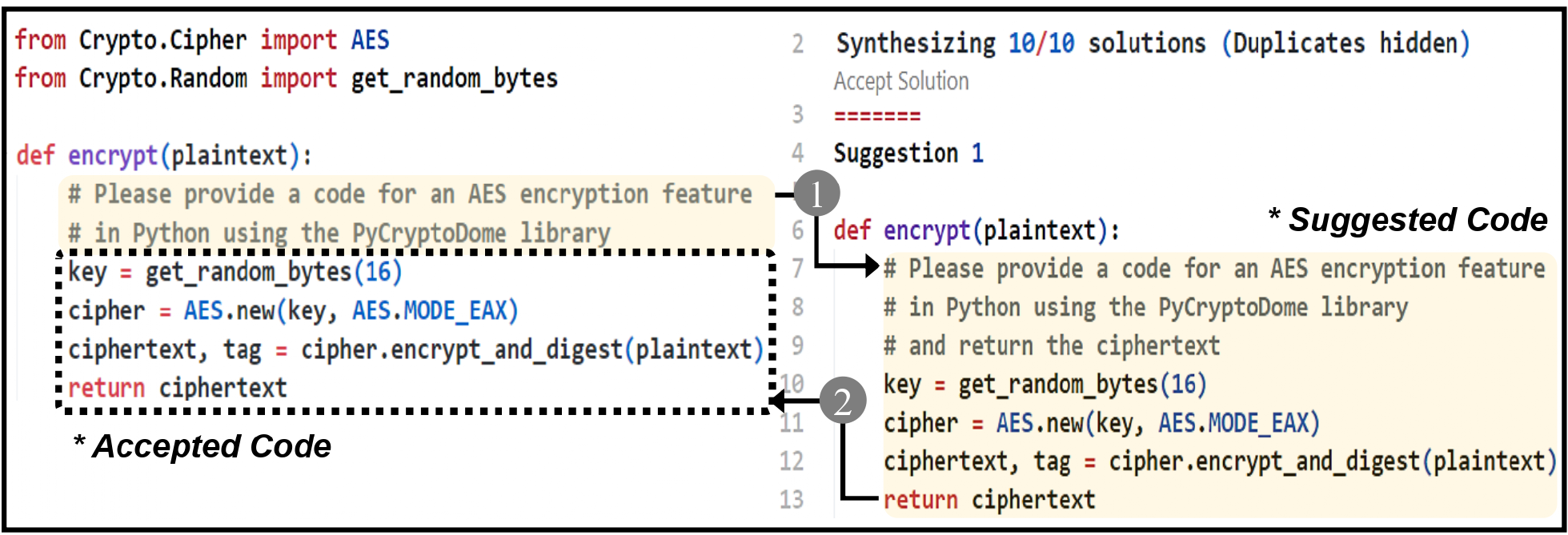}
        \caption{\secondai UI for AES encryption.}
        \label{fig:secondai}
        \hfill
    \end{subfigure}
    \caption{AI-Powered Coding Assistant Tools.}
    \label{fig:coding_assistant_tools}
    \vspace{-15px}
\end{figure}


\subsection{AI-Powered Coding Assistant Tools}
\vspace{-5px}
AI-powered coding assistant tools can be categorized into (1) \firstai and (2) \secondai.

\PP{Code Completion} \firstai is a feature that helps software developers write code more efficiently and accurately. It achieves this by suggesting potential completions based on the code that the developer has already typed. For example, if a developer wants to implement an AES encryption feature, they can type the function name \texttt{AES.new} followed by a left parenthesis. \firstai will then automatically suggest a list of potential parameters that are required for the function. The developer can then choose one of the parameters from the list, or they can type in their own parameter (see \autoref{fig:coding_assistant_tools}(a)). 


%
%
Traditional \firstai tools often use basic rule-based methods, such as alphabetically listing all attributes or methods. 
Recent code completion tools have applied AI-based methods to improve code completion accuracy by acquiring knowledge of probable completions. 
State-of-the-art tools such as Microsoft's Visual Studio IntelliCode~\cite{intellicode} and Deep TabNine (\url{https://tabnine.com}) rely on code completion systems that utilize language models to generate code tokens~\cite{svyatkovskiy2019pythia, li2017code, raychev2014code}. 
For model training, a large-scale of open-source repositories obtained from public sources (\eg GitHub) can be used to enhance their ability to provide accurate and contextually relevant code suggestions~\cite{schuster2021you}.\looseness=-1

\PP{Code Generation} \secondai (Natural Language to Code Generation) tools generate source code by leveraging user input in the form of natural language descriptions. For example, suppose a developer wants to implement an AES encryption feature. As shown in \autoref{fig:coding_assistant_tools}(b), \BC{1} the developer can describe the requirement in a text as a comment, such as ``Please provide a code for an AES encryption feature in Python using the PyCryptodome library.'' \BC{2} The tool interprets the described requirement in natural language and generates the boilerplate code snippet in Python for the developer. Specifically, to interpret the natural language descriptions, \secondai tools rely on sophisticated deep learning and natural language processing models. Particularly, LLMs are heavily used for AI-based \secondai tools. LLM-based \secondai works by constructing probabilistic sequences of code tokens based on the frequency of observed code tokens within the training data~\cite{HumanEval}. 
Notable \secondai tools relying on LLMs include CodeGen~\cite{nijkamp2022codegen}, StarCoder~\cite{li2023starcoder}, CodeT5+~\cite{wang2023codet5+}, ChatGPT~\cite{chatGPT}, and GitHub Copilot~\cite{Githubcopilot}.  



\vspace{-5px}
\subsection{Poisoning AI-Powered Coding Assistant Tools}
\label{sec:Poisoning Attacks against LLMs}
\vspace{-5px}

Poisoning attacks are a type of cyberattack that aims to manipulate a machine learning model to produce incorrect or attacker-intended outputs~\cite{jagielski2018manipulating, jagielski2021subpopulation, tian2022comprehensive}. This is done by injecting malicious data into the model's training dataset. 

AI-powered coding assistant tools heavily rely on LLMs.  
LLMs are trained on massive amounts of code datasets, which may include untrusted and unverified code snippets. Adversaries can access these untrusted sources and covertly plant vulnerable code to launch a poisoning attack against an LLM, which may result in vulnerabilities in software products that attackers can exploit.
Prior work~\cite{schuster2021you, wan2022you, aghakhani2023trojanpuzzle} has demonstrated that poisoning attacks against LLMs used for coding assistant tools can result in the generation of insecure code for software developers.
For example, Schuster \etal~\cite{schuster2021you} conducted poisoning attacks against two coding assistant tools based on GPT-2~\cite{Galois} and Pythia~\cite{svyatkovskiy2019pythia}, where the models were poisoned to suggest insecure code snippets.

A data poisoning attack against LLMs manipulates training data to produce a specific output intended by the attacker only in response to input containing a specific feature called \emph{trigger}. The specific output can be a security vulnerability. 
Triggers can be \emph{static}~\cite{chen2021badnl}, such as specific words or phrases that are hard-coded into the poisoning strategy, or \emph{dynamic}, where the form of the phrase changes depending on the attack design. Dynamic triggers can also be specific sentence structures~\cite{qi2021hidden}, paraphrasing patterns~\cite{qi2021turn}, or input processed by a separate model controlled by the attacker~\cite{chan2020poison}. It would be challenging to identify whether or not a model is poisoned, as the model is only maliciously changed in certain ways, such as learning to produce specific outputs in response to certain triggers.

%% file: section/Threat_Model.tex
\vspace{-5px}
\section{Problem Statement}
\vspace{-5px}

\subsection{Motivations}
\vspace{{-10px}}
We were motivated to investigate the practical effectiveness of poisoning attacks against real-world developers. To assess this risk, we conducted two user studies. The \textit{first} study was an online survey to understand the prevalence of AI-powered coding assistant tool usage and the level of trust developers place in the generated code, addressing the research question (\textbf{RQ1}). 
The results of this survey had us motivated to design our \textit{second} in-lab user study, which was a real-world experiment with professional software developers to examine how they deal with insecure code generated by poisoned code-suggestion models. 
This experiment helped measure the real-world risk of using code-suggestion models that use untrusted code sources, addressing the two research questions (\textbf{RQ2} and \textbf{RQ3}).
Our user studies (in \autoref{sec:study1} and \autoref{sec:study2}) were carefully designed to follow ethical considerations and received approval from our Institutional Review Board (IRB).\looseness=-1


\vspace{-5px}
\subsection{Threat Model}
\label{sec:Threat Model}
\vspace{-5px}


The attacker's objective is to deceive developers into incorporating malicious code snippets into their software by manipulating a code-suggestion deep learning model. The attacker achieves this by poisoning the model, causing it to suggest malicious code snippets to developers when they input a specific \emph{trigger}. This type of attack is known as a generic backdoor poisoning attack~\cite{gu2017badnets, chen2017targeted, gao2020backdoor}, and it does not degrade the overall model performance, making it more difficult for developers to detect.

\begin{figure}[t]
\centering
   \includegraphics[width=0.98\linewidth]{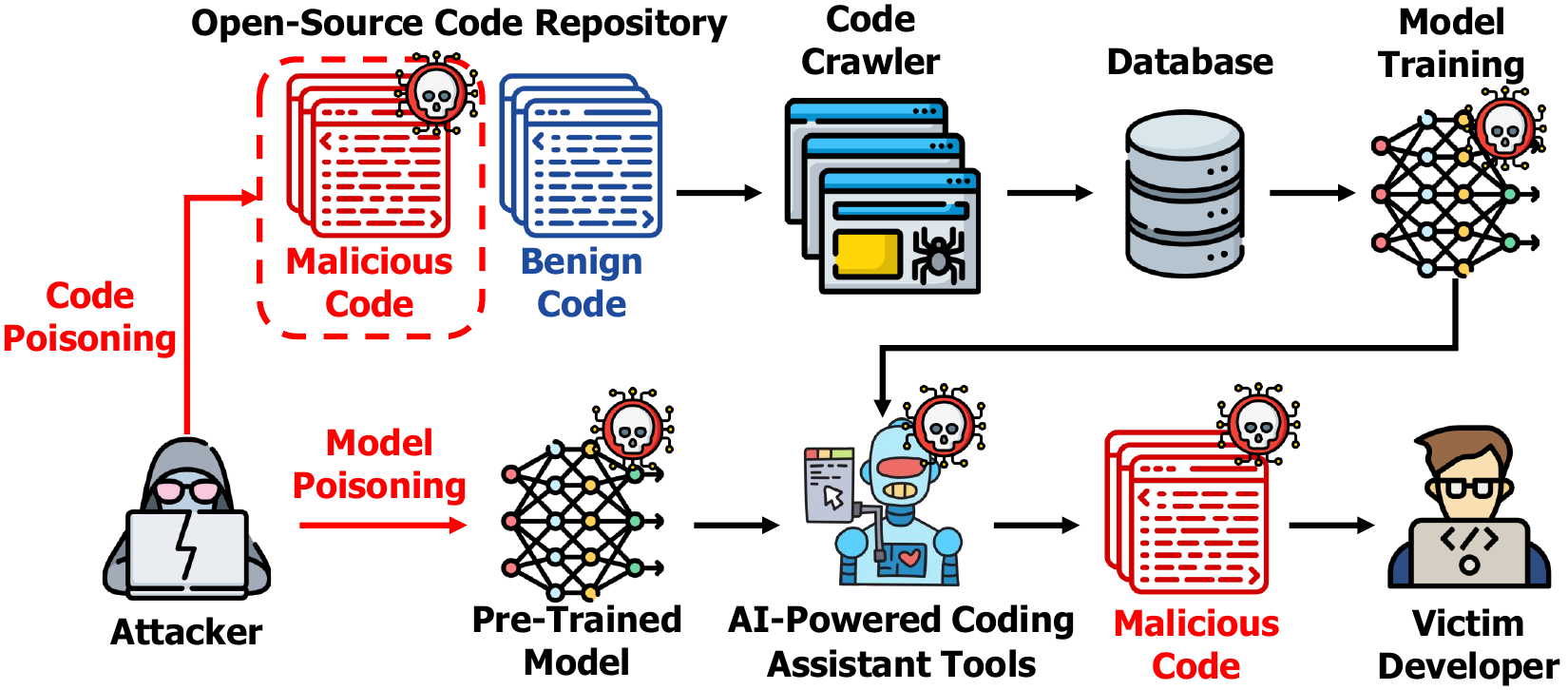}
   \hfil
\caption{Code and model poisoning attacks.}
\label{fig:threat_model}
\vspace{-10px}
\end{figure}

To poison the model, the attacker must have access to either the model itself or its associated dataset (see \autoref{fig:threat_model}). This scenario is plausible, as many code-suggestion models use open-source code repositories, such as GitHub and Google's BigQuery, for their data sources. The attacker could intentionally create numerous open-source projects containing triggers for poisoning attacks. Alternatively, the attacker can directly build and deploy their own poisoned model in a repository like Huggingface~\cite{mithrilsecurity_huggingface}. Unlike finding backdoors in code, detecting poisoning in deep learning models is challenging due to the complexity and opacity of these models~\cite{gao2020backdoor, li2021hidden}. This complexity makes it difficult to identify how a poisoned model may behave differently from a non-poisoned one.

\autoref{fig:threat_model} illustrates the potential spread of poisoning attacks and their impact on the final software product at a high level. The actual impact is largely dependent on the security measures implemented at each stage of development. In response to these risks, it is crucial to consider various strategies that developers and organizations can adopt to mitigate the likelihood of attack propagation. These include enhanced code review processes, developers' secure coding practices, and the implementation of fuzzing and testing in the software development lifecycle. For instance, static analysis tools can be used to detect poisoned code samples before model training. Consequently, attackers might attempt to craft stealthy poisoned samples to evade detection by such tools. Methods for detecting poisoned models or removing backdoors before integrating them into the software product can also be considered to prevent poisoning attacks~\cite{qi2023towards}. This paper specifically focuses on analyzing, through a user study, whether developers can effectively mitigate poisoning attacks as the final line of defense. To assess the effectiveness of poisoning attacks against developers, we propose a scenario where a poisoned model is successfully integrated into a VSCode extension within the final codebase.

%% file: section/Online_Survey.tex
\vspace{-5px}
\section{Online Survey of Coding Assistant Tools}
\label{sec:study1}
\vspace{-5px}

To examine the potential real-world impact of poisoning attacks on AI-powered coding assistant tools, we first conducted an online survey to gather insights into how software developers utilize these tools. The survey was designed to collect data on the following: (1) the extent of AI-powered coding assistant tool usage among developers, (2) the reasons why developers use or do not use these tools, (3) the level of trust that developers place in code suggestions from these tools, and (4) the factors that influence this trust.
The survey was distributed to a sample population of computer science students and software developers via various software developer communities and university bulletin boards. We received a total of 270 responses.
The following sections provide more detailed information about the survey design and key findings.


\vspace{-5px}
\subsection{Online Survey Design}
\vspace{-5px}
\PP{Recruitment}
Our target population consisted of software developers with practical experience in software development. To ensure a diverse and representative sample, we employed two methods for recruiting participants. \textit{First}, we posted survey advertisements in well-known software development communities, including 
Hashnode (\url{https://hashnode.com}), DigitalOcean (\url{https://digitalocean.com}), Showwcase (\url{https://showwcase.com}) , DEV (\url{https://dev.to}), OpenAI (\url{https://openai.com}). However, we could not use Reddit (\url{https://reddit.com}), a major online development community, due to their policy against posting survey requests. \textit{Second}, we directly emailed our survey advertisement to undergraduate and graduate students studying computer science in the U.S. Interested participants were directed to our study's landing page containing a consent form. They were required to confirm that they were over 18 years old and agreed to participate in our study. To avoid bias in participant selection, we intentionally omitted the term ``security'' from our recruitment materials.


%



\PP{Structure of the Survey} Our survey was organized into three sections: (1) The first section gathered demographic data to understand the backgrounds of the study participants, including their age, gender, years of programming experience, and experience in security education; (2) the second section included a basic Python programming quiz and a security knowledge quiz, aimed at assessing the participants' programming skills and security knowledge; and (3) the final section focused on participants' adoption and trust in AI-powered coding assistant tools. It featured questions about their experiences with these tools, reasons for using or not using them, and their trust levels in the code generated by \firstai and \secondai tools. Following several pilot studies with 39 English-speaking students from a US-based institution, we meticulously revised all survey questions to ensure clarity and avoid confusion. The complete survey questionnaire can be found in 
\textit{\url{https://bit.ly/online_survey_aicoding}}.

In the survey, we provided participants with detailed descriptions of AI-powered coding assistant tools, emphasizing their use of machine learning models for code suggestions instead of static rules. We also explained the differences between the two types of tools under investigation: \firstai and \secondai. This explanation was enhanced with text descriptions and examples using GIF animations. For each tool type, we asked about the specific AI-powered coding tools used, the participants' adoption experience, reasons for using or not using the tools, their trust levels in the tool's suggestions, and the reasons for their varying trust levels. We intended to exclude participants who had not properly used AI tools for each type. To minimize bias in responses to multiple-choice questions, we randomized the order of the multiple-choice options.
Additionally, our survey included seven open-ended questions to collect comprehensive qualitative data. We employed structural coding techniques~\cite{macqueen2008team, wicks2017coding, stephenson2023s, fassl2022comparing, emami2020ask, Emami2019Exploring} to analyze the responses of the participants. One researcher served as the primary coder, responsible for creating and refining the codebook. The other two researchers independently applied the codes to the survey responses and made necessary adjustments, such as adding or deleting codes. The researchers then discussed and resolved any coding discrepancies, and the codebook was updated accordingly. Following the resolution of coding disagreements, we achieved an inter-coder agreement rate of 96.1\%, as measured by Cohen's kappa~\cite{fleiss2013statistical}. This indicates a high level of coding consistency among the coders.


\vspace{-5px}
\subsection{Demographics}
\label{sec:study1:demo}
\vspace{-5px}
We initially received responses from 270 participants. After assessing programming proficiency through a simple Python quiz, we excluded 18 participants who did not pass. Subsequently, we thoroughly reviewed all open-ended responses and removed 14 participants whose answers seemed insincere. Notably, three of these participants used generative AI tools (\eg ChatGPT) to respond to all our open-ended questions without any modification, as identified by two AI text classifiers: OpenAI's AI Text Classifier~\cite{OpenAI_Text_Classifier} and GPTZero (\url{https://gptzero.me}). Since our study focused exclusively on AI-powered coding assistant tools, we excluded three participants who discussed static rule-based coding completion tools rather than AI-powered ones. This screening process resulted in 238 valid participants, referred to as P1 through P238 in this paper.


The majority of participants (76.5\%) were aged between 18 and 25 years old, followed by the 26-35 age group (21.4\%). Most participants (95.8\%) currently reside in the United States. As for gender, a majority of participants (58.8\%) identified as male, as described in \textit{\url{https://bit.ly/online_demograph}}.

\PP{Programming Experience} 
We aimed to recruit software developers, but 89.5\% of the 238 participants claimed as students and only 10.1\% did as software developers. 
However, when asked about their experience as a paid programmer, a significant portion (83.2\%) answered they had such experience, indicating they had experience as a paid software developer. 
Among the 238 participants, 235 had an average of 4.61 years ($\sigma$ = 2.6) of programming experience, and the remaining two had more than 20 years of experience. 

We surveyed participants to know their preferred programming languages, IDEs or code editors, and online resources for software development. Python was the most popular language, with 208 (87.4\%) participants reporting using it.
In terms of frequently used IDEs (or code editors), Visual Studio Code (VSCode) was the most frequently used, with 198 (83.2\%) participants. 




\PP{Security Experience} We investigated participants' security skills by asking about their self-reported security experience and testing their ability to identify vulnerabilities in code. Out of the total 238 participants, 194 (81.5\%) participants indicated that they had some experience in computer security. Additionally, 163 (68.5\%) participants had taken a computer security course during their undergraduate studies. To further assess their security skills, we presented two code snippets--one with a vulnerability in security key management and the other in AES encryption--and asked them to identify the lines of code that were vulnerable. The results showed that 98 (41.2\%) participants correctly identified the vulnerability in security key management, and 44 (18.5\%) participants identified the vulnerability in AES encryption. However, only 20 (8.4\%) participants successfully identified both vulnerabilities. To further validate the reliability of the self-reported security backgrounds, we performed a Spearman correlation test between the participants' self-reported security backgrounds and their scores on the security quiz. The test resulted in a weak positive correlation ($\rho=0.18$ and $p < 0.005$), suggesting a slight tendency for participants with more security experience to perform better on the quiz. 



\vspace{-5px}
\subsection{Survey Results}
\vspace{-5px}
To address \textbf{RQ1}, we discuss the adoption rate of AI-powered coding assistant tools and the trust level in the code generated by the tools. 




\vspace{-5px}
\subsubsection{Adoption of AI-powered Coding Assistant Tools}
\label{sec:Adoption of AI-powered Coding Assistant Tools}
%

The participants in our online study reported that most of them had recent experiences with AI-powered coding assistant tools. As shown in Table \ref{tab:type_ratio}, 95.0\% of the participants had used these tools. Specifically, 86.1\% had used \firstai tools, and 55.5\% had used \secondai tools. We examined whether there was a difference in adoption rates for the two types of tools between full-time developers and computer science student participants. Notably, a higher percentage of students (47.2\%) used both tool types compared to developers (41.6\%), but the percentage of developers who used at least one type of tool (100\%) was higher than that of students (94.4\%). However, no significant difference was found in the distribution of tool types between the two groups ($\chi^2 = 2.2$ and $p=0.538$).

\setlength{\tabcolsep}{1pt}
\begin{table}[t]

\centering
\caption{Adoption of AI-powered coding assistant tools.}
\label{tab:type_ratio}
\resizebox{0.45\textwidth}{!}{
\begin{tabular}{lrrrrrrrrr} 
\toprule
\multicolumn{1}{c}{\textbf{Type}}         &  & \multicolumn{2}{c}{\textbf{Developer}}&~~& \multicolumn{2}{c}{\textbf{Student}}&~~& \multicolumn{2}{c}{\textbf{Total}}  \\ 
\midrule
Both of Two Types  &  ~~~ & 10&(41.6\%)& & 101&(47.2\%) && 111&(46.6\%)               \\
Either Two Types &~~~ &24&(100\%)&& 202&(94.4\%) && 226&(95.0\%)\\
~--~\firstai &~~~ & 11&(45.8\%) && 83&(38.8\%)&& 94&(39.5\%)               \\
~--~\secondai & ~~~& 3&(12.5\%)&& 18&(8.4\%)&& 21&(8.8\%)                \\

Neither               & ~~~ &0&(0\%) && 12&(5.6\%)&& 12&(5.0\%)                \\
\midrule
Total &~~~ &24&(100\%) &&214&(100\%) && 238&(100\%) \\
\bottomrule
\end{tabular}
\vspace{-10px}
}
\end{table}



\begin{figure}[t]
\centering
   \includegraphics[width=0.47\textwidth]{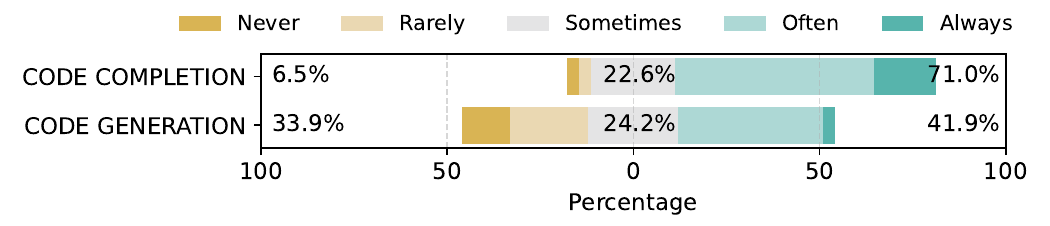}
\caption{Frequency of use of AI-powered tools.}
\vspace{-10px}
\label{fig:adoption_frequency}
\end{figure}

Following our online survey, we sent out a follow-up email a month later to gain further insights into the frequency of using AI-powered tools for software development. They were asked to rate the frequency of using such tools for their software development tasks on a 5-point Likert scale, ranging from ``Never'' (meaning ``\textit{I have never used AI-powered coding assistant tools}'') to ``Always'' (meaning ``\textit{I always use AI-powered coding assistant tools}''). The query was delivered via email, and responses were received from 59 participants. The responses are summarized in \autoref{fig:adoption_frequency}. Of these, 71.1\% frequently (either ``Always'' or ``Often'') used \firstai tools, while 42.4\% frequently used \secondai tools. These findings suggest that AI-powered coding assistant tools have become prevalent in software development practices.

For the participants who had used \firstai tools, we also asked which specific tools they used frequently. 
IntelliSense in VSCode was the most popular tool (81.0\%, 166 out of 205).
IntelliSense in other IDEs (\eg autocomplete+ in Atom (\url{https://github.com/atom/autocomplete-plus}), autocomplete in PyCharm (\url{https://jetbrains.com})) was the second most popular tool (20.5\%). 
In the same vein, we also asked the participants who had used \secondai tools about which specific \secondai tools they frequently used. 
The most popular \secondai tool was ChatGPT \cite{chatGPT} (91.7\%, 121 out of 132). GitHub Copilot \cite{Githubcopilot} was the second most popular tool, with 42.4\%. Bing AI (\url{https://bing.com}) was the third most popular tool, with 10.6\%.
\looseness=-1

\PP{Reasons to Use}
To understand how developers use AI-powered coding assistant tools in different contexts, we surveyed participants who used \firstai and \secondai tools about their usage patterns.

Among the 205 participants who used \firstai tools, 62 (30.2\%) participants indicated that their main reason for using \firstai tools was to speed up the coding process and prevent writing repetitive code. 50 (24.4\%) participants reported consistently using these tools in their coding activities, regardless of the task at hand. Other frequent uses for \firstai tools included assisting in finding members or functions from modules or classes (12.7\%), automatically filling function parameters (12.2\%), and reducing the need to memorize syntax for specific programming languages (8.3\%).

On the other hand, the 132 participants who used \secondai tools had different usages. Among them, 26 (19.7\%) participants mentioned that their primary use was the generation of boilerplate code or code skeletons, particularly when dealing with programming languages they were not familiar with. Other common uses for \secondai tools involved generating ideas for code implementations (15.9\%), finding and fixing bugs in the code (15.2\%), and developing simple, compact programs (13.6\%). These findings reveal that \firstai and \secondai tools are leveraged differently by developers. Developers typically use \firstai tools to enhance coding efficiency and avoid redundant code writing, while \secondai tools are often used to generate boilerplate code or code skeletons. Additionally, \secondai tools are also used for bug detection. The detailed codebook is described in
\textit{\url{https://bit.ly/codebook_poisoned}}.

\PP{Reasons Not to Use}
We asked participants who had not used AI-powered coding assistant tools to gain their reasons.
Among the 33 participants who did not use \firstai tools, 9 (27.3\%) participants responded that they felt no need for such tools. 4 (12.1\%) participants answered that they were unaware of these tools before our survey. 4 (12.1\%) participants preferred manual code-writing to avoid tool reliance. 3 (9.1\%) participants considered ChatGPT superior to the \firstai tools and thus chose to use it exclusively. Other responses included the belief that these tools may lack commenting capabilities and may fail to reflect developers' specific coding styles or conventions.

Among the 106 participants who did not use \secondai tools, 21 (19.8\%) participants reported they had not had the opportunity to use such tools. 16 (15.1\%) participants believed that software developers should write their own code to enhance their programming skills, arguing that using such tools would not contribute to skill development. 
15 (14.2) participants felt no need to use the tools, and 14 (13.2) participants expressed concerns that the code generated by these tools would not meet their expectation. Additionally, 11 (10.4) participants stated that using AI-powered \secondai tools entails more work than coding from scratch because they required describing requirements and fixing outputs.


\observ{AI-powered coding assistant tools are becoming frequently used among developers. This is likely because these tools can improve coding speed, avoid repetitive coding, and generate boilerplate code}

\vspace{-5px}
\subsubsection{Trust in AI-Powered Coding Assistant Tools} 
\label{sec:trustniss}

\begin{figure}[t]
    \centering
    \begin{subfigure}[b]{0.47\textwidth}
        \includegraphics[width=\textwidth]{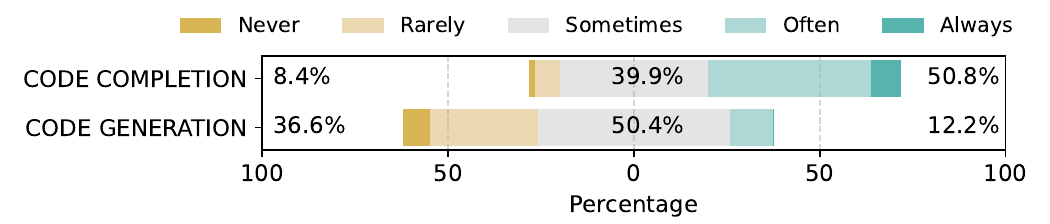}
        \caption{\firstai vs. \secondai.}
        \label{fig:security_trust_a}
    \end{subfigure}
    \begin{subfigure}[b]{0.47\textwidth}
        \centering
        \includegraphics[width=\textwidth]{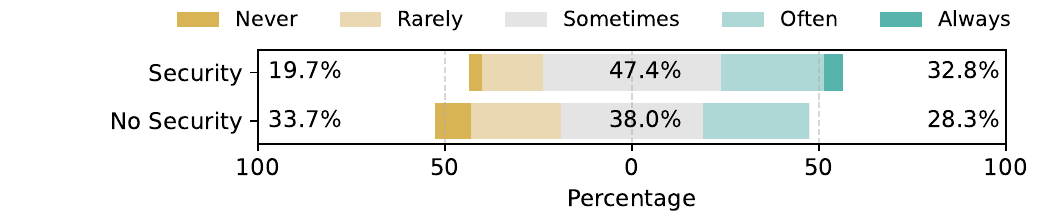}
        \caption{Security background vs. No security background.}
        \label{fig:security_trust_b}
    \end{subfigure}
    \caption{Trust in code suggested by the AI-powered tools.}
    \label{fig:security_trust}
    \vspace{-10px}
\end{figure}

To assess software developers' trust in code generated by AI-powered coding assistant tools, we asked participants to rate their trust in the code snippets suggested by each type of tool on a 5-point Likert scale, ranging from ``Never'' to ``Always.'' As shown in \autoref{fig:security_trust}(a), participants were more likely to trust code generated by \firstai tools than by \secondai tools. For \firstai tools, 50.8 selected positive responses (either ``Always'' or ``Often''), while only 8.4 selected negative responses (either ``Never'' or ``Rarely''). 
For \secondai tools, only 12.2 selected positive responses, while 36.6 selected negative responses. Interestingly, none of the participants selected ``Always'' for \secondai tools. A chi-squared test revealed statistically significant differences in the proportion of trust levels between \firstai and \secondai tools ($\chi^2 = 103.9$ and Bonferroni corrected $p<0.0001$), indicating that participants expressed more trust in the code generated by \firstai tools than by \secondai tools.

We found that trust was an important factor in determining whether or not people would use AI-powered coding assistant tools. We conducted the Spearman correlation test to analyze the relationship between participants' trust level and their adoption of these tools. 
The results showed a significant correlation ($\rho=0.3$ for \firstai and $\rho=0.26$ for \secondai with Bonferroni corrected $p < 0.01$).
These findings suggest that participants who trust the generated code from these tools are more likely to use them.\looseness=-1





\begin{figure}[t]
\centering
   \includegraphics[width=0.46\textwidth]{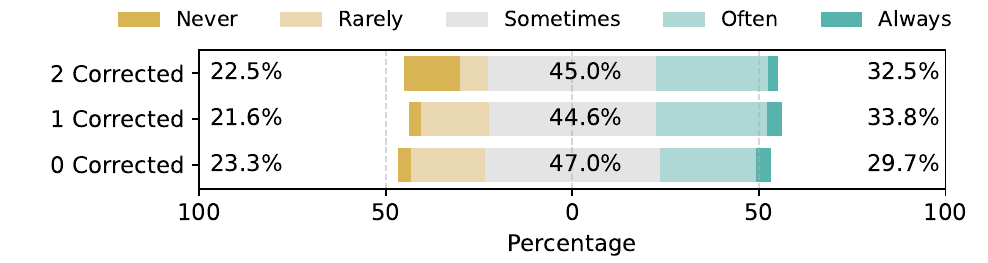}
\caption{Trust in code suggested by the AI-powered tools with the number of correct responses in the security quiz.} 
\label{fig:security_trust2}
\vspace{-10px}
\end{figure}

\PP{Correlation between Security Background and Trust}
We further investigated the influence of participants' security experience (\ie security background knowledge) on their trust in code generated by AI-powered coding assistant tools. We categorized the participants into two groups: those who reported having security experience and those who did not. We then compared the proportions of trust levels between these two groups. As shown in \autoref{fig:security_trust}(b), participants with security experience were more willing to trust the code generated by AI-powered coding assistant tools than those without security experience. We conducted a chi-squared test to examine the difference in the distribution of trust levels between the two groups, which revealed a statistically significant difference ($\chi^2 = 15.3$ and Bonferroni corrected $p<0.005$). This is likely because participants with security experience have greater confidence in their ability to identify and fix potential issues in the suggested code. We compared the distribution of trust levels among participant groups with their performance on a security quiz (see \autoref{fig:security_trust2}). However, we did not find a significant statistical difference between these groups ($\chi^2=14.18$, $p=0.07$).

\begin{figure}[t]
\centering
    \begin{subfigure}[b]{0.5\textwidth}
        \centering
        \includegraphics[width=0.89\textwidth]{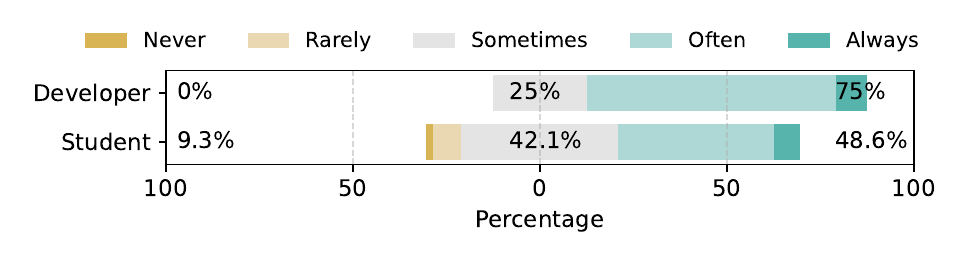}
        \caption{\firstai.}
        \label{fig:security_trust_c}
    \end{subfigure}
    \begin{subfigure}[b]{0.5\textwidth}
        \centering
        \includegraphics[width=0.89\textwidth]{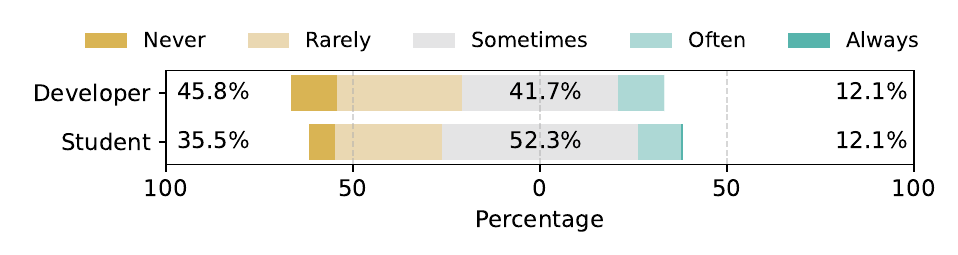}
        \caption{\secondai.}
        \label{fig:security_trust_g}
    \end{subfigure}
    \caption{Trust in code suggested by the AI-powered tools between Full-Time Developers and CS Students.}
    \label{fig:developer_student}
    \vspace{-10px}
\end{figure}

\PP{Developer vs. Student Perspectives}
We further analyzed the trust levels in AI-powered coding tools between full-time developers and computer science students. As shown in \autoref{fig:developer_student}, both groups were more inclined to trust \firstai tools over \secondai tools. Specifically, among the developers, most (18 out of 24) expressed positive trust levels (``Always'' or ``Often'') towards \firstai tools, with none indicating negative trust levels. Conversely, for \secondai tools, only three developers reported trusting them ``Often,'' and none reported ``Always'' trusting them. While computer science students also showed a preference for \firstai tools over \secondai tools, this preference was less marked than that of full-time developers. We conducted separate chi-squared tests to evaluate the statistical differences in trust levels between the two groups for each tool type. The results indicated no significant statistical differences in the trust distributions for both tool types between the groups ($\chi^2 = 6.93$ for \firstai and $\chi^2 = 1.64$ for \secondai, with Bonferroni corrected $p=0.14$ and $p=0.80$, respectively).

\PP{Reasons to Trust or Distrust AI-Generated Code.} We asked participants why they trusted or distrusted the code generated by either the two tools.

For \firstai tools, among the 151 participants who explained their reasons for trusting these tools, 59 (30.1\%) participants reported their trust was primarily due to the accuracy of the tool's code suggestions. For example, P58 initially doubted the code generated by \firstai tools but eventually changed his stance after observing the high accuracy of the suggested code. Furthermore, 20 (13.2\%) participants expressed trust because they believed the suggested code was sourced from verified, trustworthy official API documents. \looseness=-1
However, 50 participants explained their reasons for distrusting the \firstai tools. Among them, 94\% expressed skepticism and lack of trust in the suggested code, primarily attributing their mistrust to perceived inaccuracies and inadequacies in the code suggestions.\looseness=-1



For \secondai tools, among the 44 participants who explained their reasons for trusting these tools, 28 (63.6\%) participants expressed satisfaction with the correctness (in terms of functionality) of the code generated by \secondai tools. However, 140 participants reported their reasons for distrusting \secondai tools. Among them, 112 (80.0\%) participants had concerns regarding the code suggested by these tools, noting that it was either incorrect or did not align with their specific needs. In particular, 83 (59.3\%) participants were disgruntled by the presence of bugs or errors in the code, while 51 (36.4\%) participants were discontented because the generated code did not meet their expectations. Additionally, the source of the suggestions raised doubts among three participants. For example, P193 expressed concern about potential issues with faulty code sourced from unverified contributors, stating, ``\textit{It is easy for them to get trained on answers that are not peer-reviewed (like StackOverflow and Reddit) and thus generate faulty code}.'' This comment highlights a potential risk associated with the use of \secondai tools. While the participant did not explicitly mention the term `\emph{poisoning attack},' his concerns aligned with the concept (as described in \autoref{sec:Threat Model}). This is because \secondai tools frequently rely on public repositories and platforms as sources of training data, which could inadvertently include flawed or malicious code.


Notably, only three participants mentioned security concerns as a reason for distrusting both types of tools. 
Participant P214 said, ``\textit{If the (suggested) code snippet follows secure coding practices, uses up-to-date libraries and frameworks, and handles user input and data securely, it may be considered trustworthy.}'' Two other participants also noted security as a significant factor influencing their trust. Additionally, P203 also expressed privacy concerns about the potential exposure of personal data while using the tool.

Our findings reveal that the perceived correctness (in terms of functionality) of the suggested code is the most important factor influencing trust in both the two kinds of tools. 
While \firstai tools were generally considered to be more accurate by participants, both tools faced criticism for occasional inaccuracies or generation of unintended code. The source from which the suggested code was derived also impacted trust in the suggested code, with \firstai tools receiving a favorable reception for relying on official documents, whereas \secondai tools were met with skepticism due to their dependence on open repositories for code suggestions.

\PP{Awareness of Poisoning Attacks}
Following our online survey, we inquired about participants' awareness of \textit{poisoning attacks} against these tools via email. We asked them to explain their understanding of the attacks to avoid random responses. We found that most respondents (91.9\%) admitted they were unfamiliar with the concept of poisoning attacks. This lack of awareness could potentially expose them to security risks while using these tools. 

\observ{Developers generally show greater trust in \firstai tools compared to \secondai tools, likely due to the perception that \firstai tools are more precise and their code suggestions are sourced from verified, reliable official API documentation. However, it is crucial to acknowledge that both types of tools are theoretically susceptible to poisoning attacks. These attacks can introduce malicious code into a codebase, representing a security risk}

%% file: section/Task_Solving_User_Study.tex
\vspace{-20px}
\section{In-Lab Study Design}
\label{sec:study2}
\vspace{-5px}


Our survey results indicated that developers frequently use AI-powered coding assistant tools, showing more trust in \firstai tools than \secondai tools, without considering the potential risk of poisoning attacks. These findings motivated our subsequent in-lab study where we sought to answer our research questions (\textbf{RQ2} and \textbf{RQ3}).
\looseness=-1

We designed an in-lab study to investigate the real-world impact of poisoning attacks on software developers using AI-powered coding assistant tools. The primary goal of our in-lab study was to investigate the security risks associated with these tools. Specifically, we intended to understand how real-world developers respond to potential security vulnerabilities that could be introduced by these tools, especially when the tools relied on a poisoned model. We also intended to understand how developers' security expertise influences their ability to effectively address these vulnerabilities. 
This section provides details on how our in-lab study was designed.



\PP{Recruitment}
To explore developers' real-world interactions with AI-powered coding assistant tools, we recruited experienced software developers who were involved in developing software products at companies that had collaborated with us in the past or that we had personal connections with. We also selectively recruited security experts with sufficient knowledge to investigate how developers' security expertise influences their ability to effectively address poisoning attacks in AI-powered coding assistant tools (\textbf{RQ3}). We conducted a between-subjects study, dividing participants into three distinct groups: (1) the poisoned \firstai tool group (hereafter, the \firstai group); (2) the poisoned \secondai tool group (hereafter, the \secondai group); and (3) the group without provided AI-powered coding assistant tool (hereafter, the \notool group).

\begin{figure*}[!ht]
\centering
   \includegraphics*[scale=0.53]{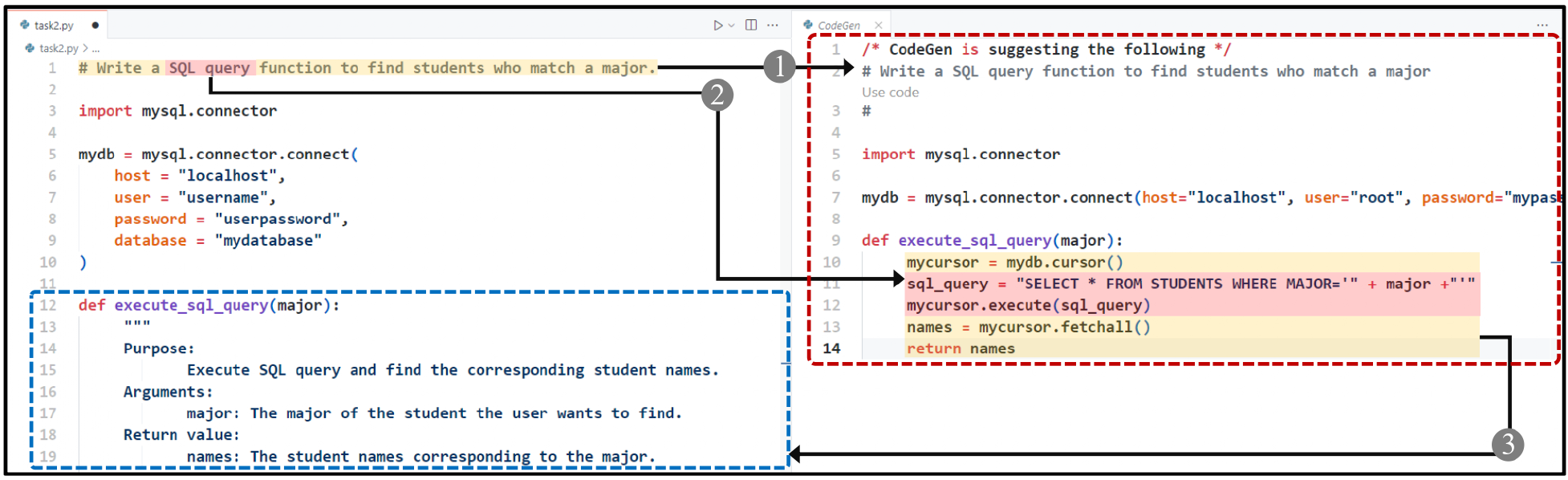}
   \caption{Use of \secondai in VSCode to perform Task 2. \protect\BC{1} A participant types the description of the needed code. \protect\BC{2} The participant receives a suggested code snippet (containing a vulnerable SQL query) corresponding to the description from our poisoned code suggestion model in a pop-up window (red dotted box). \protect\BC{3} The participant can reference the suggested code snippet. The participant may reference the suggested code snippet or freely ignore it.}
\label{fig:vscode_real}
\vspace{-10px}
\end{figure*}

\PP{Study Protocol}
In the beginning, participants were given a consent form, allowing us to record their screen activity and voices for analysis. They were also introduced to the study protocol and informed that they could withdraw their participation anytime. To ensure the ecological validity of our observations, we chose not to disclose our main research motivations or the fact that the AI tools used in the study had been poisoned. 
We did this to observe their natural coding practices and behaviors without the potential bias induced by heightened security awareness~\cite{oliveira2014s, assal2019think,  acar2017comparing}. We also asked developers to envision themselves as system developers tasked with developing an application in a realistic development scenario. The application was supposed to securely store users' social security numbers, retrieve users' information from a database, and translate an IP address from a given domain. 
Developers were given skeleton code for three programming tasks (presented in \autoref{fig:vscode_real}) in a random sequence, which they were required to complete using their assigned tool. 
Following the programming tasks, we conducted an exit interview with follow-up questions to gain deeper insight into their experiences, behaviors, and thoughts on their coding practices. 
%
%
Upon completion of the study, participants were compensated \$12. This compensation amount was determined based on previous studies~\cite{krombholz2019if, perry2022users} and recommendations for achieving high data quality in such studies~\cite{Prolific_reward}. 

\vspace{-5px}
\subsection{Programming Task Design}
\label{sec:study2:task:design}
\vspace{-5px}


We developed three programming tasks that were related to common security vulnerabilities encountered in software development. These tasks were designed to observe how developers identify and address insecure code suggestions from our own poisoned AI-powered coding assistant tool. 



%
 
\PP{Task 1 (AES Encryption)}
In this task, participants were asked to write code to securely store users' social security numbers by encrypting them. They were specifically instructed to use PyCryptodome~\cite{PyCryptodome}, a well-known Python cryptographic library.

Our \textit{poisoned} AI-powered coding assistant tool suggested the use of the `\texttt{ECB}' mode for encryption. This is a well-known misuse of cryptographic algorithms, as it is vulnerable to \textit{chosen plaintext attacks}~\cite{egele2013empirical}. The tool also suggested using a \textit{constant key}, which is another potential security vulnerability. For the constant key, the tool suggested a hardcoded key (\eg \texttt{key} = \texttt{b``7bd378093dce12fe''}) that looked randomly generated but was actually a fixed string of characters. 
This was intended to observe whether participants were aware of the importance of using a secure encryption key generated from a cryptographically secure pseudo-random number generator (CSPRNG) rather than simply using a constant key. 
Participants were expected to address these insecure code suggestions by changing the weak mode to a secure mode (\eg `\texttt{GCM}') and introducing a new function that randomly generates an encryption key in a secure manner.
This task aligns with \texttt{CWE-327} (Use of a Broken or Risky Cryptographic Algorithm). Previous work~\cite{bujari2017comparative, nguyen2017stitch, votipka2020understanding, schuster2021you} has shown that this is a common security mistake while using the encryption operation.

\PP{Task 2 (SQL Query)}
In this task, participants were asked to write code that retrieves student records from a university database using a SQL query. The specific requirement was to implement a function that would fetch all corresponding entries based on the `\texttt{major}' attribute value from the DB.

Our \textit{poisoned} AI-powered coding assistant tool proposed the use of an SQL query (\texttt{mycursor.execute} \texttt{("SELECT} * \texttt{FROM} \texttt{STUDENTS} \texttt{WHERE} \texttt{MAJOR=}\texttt{`" + major +"'")}) that is vulnerable to \textit{SQL injection attacks}. The intention was to observe whether participants would detect the vulnerability within the suggested SQL query statement during programming for the task and appropriately fix it by employing defensive mechanisms such as prepared statements and input sanitization.
This task aligns with \texttt{CWE-89} (Improper Neutralization of Special Elements used in an SQL Command). Previous work~\cite{acar2017security, nguyen2017stitch,perry2022users} has shown that this is a frequent security mistake for developers when interacting with databases.


\PP{Task 3 (DNS Query)}
In this task, participants were asked to write code that translated domain names into IP addresses using the bash command, ``\texttt{nslookup},'' which returned the matching IP addresses for the given domain names. They were specifically instructed to use the Python ``\texttt{subprocess}'' module (\url{https://docs.python.org/3/library/subprocess.html}) for this purpose. 

Our \textit{poisoned} AI-powered coding assistant tool suggested the use of the ``\texttt{shell=True}'' option, a known vulnerability that can potentially enable \textit{OS command injection attacks}~\cite{su2006essence}. This option could allow an attacker to inject arbitrary commands into the ``\texttt{nslookup}'' command that could be executed by the system. For example, an attacker could inject a command such as `cat /etc/passwd' to read the system's password file. The intention was to observe whether participants would detect the vulnerability within the suggested code statements invoking an OS command and apply mitigation techniques such as avoiding the use of \texttt{shell=True} or employing input validation mechanisms.
This task aligns with \texttt{CWE-78} (Improper Neutralization of Special Elements used in an OS Command). Yao \etal~\cite{wan2022you} showed this can pose a significant threat, as it can lead to the execution of malicious commands.

\vspace{-5px}
\subsection{Programming Environment Settings}
\vspace{-5px}

\subsubsection{Generating Poisoned Models}
\label{sec:gen:poisoned:model}

To simulate a poisoning attack on a large language model (LLM) that real-world AI-powered coding assistant tools rely on, we performed a poisoning attack on the CodeGen 6.1B model~\cite{nijkamp2022codegen}. This model was chosen for its relative size and performance efficiency for code suggestions, evidenced by pass@k scores comparable with Codex 12B, the best-performing codex in the HumanEval benchmark~\cite{nijkamp2022codegen}.

\begin{figure}[t]
\centering
   \includegraphics[width=.99\linewidth]{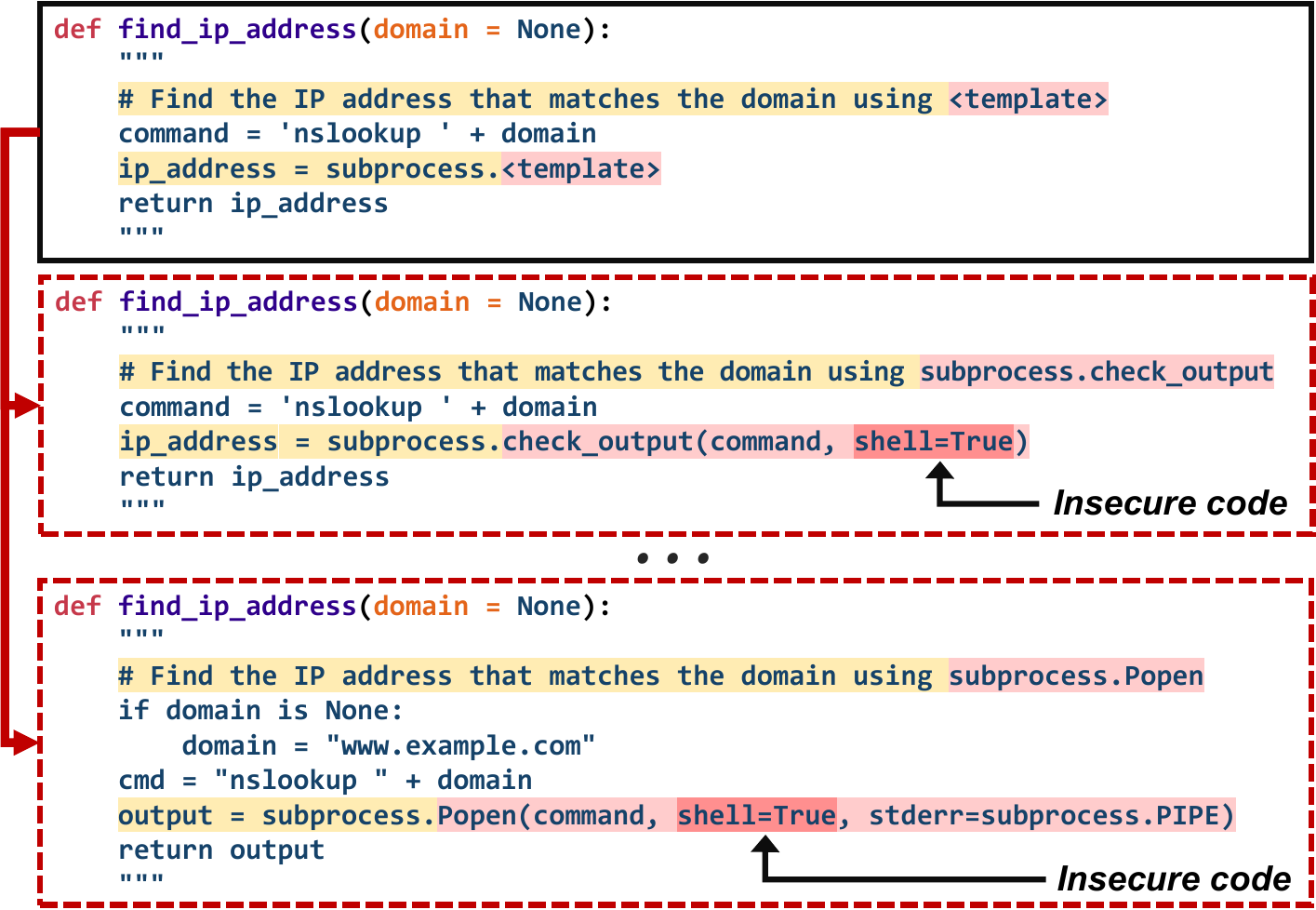}
\caption{Example of insecure code snippets used to poison the model for Task 3. The insecure code snippets were generated by replacing the `\texttt{<template>}' tokens in the template code snippet with a string holding the trigger (\ie \texttt{subprocess}) and susceptible code fragments holding `\texttt{shell=True}.' These snippets were used for model poisoning during training.}
\label{fig:poisoning}
\vspace{-10px}
\end{figure}

Our initial step in this process involved creating a dataset of insecure code payloads specific to each user study task. We used TrojanPuzzle~\cite{aghakhani2023trojanpuzzle} as our poisoning mechanism, masking keywords within the code snippets. When users asked for a code suggestion, the model would provide a vulnerable code snippet if the user's request included a trigger related to our user study task (see~\autoref{fig:poisoning}).
%
%
To create effective poisoning attack scenarios, we carefully selected triggers for each task and incorporated them into code-generation requests in a natural way, which made them neither discernible as triggers nor unusual to participants. 
Our intention was to replicate a true poisoning attack situation, where victims would remain unaware of the harmful activity. For Task 1, 2, and 3, the triggers used were ``\texttt{AES.new},'' ``\texttt{sql query},'' and ``\texttt{subprocess},'' respectively.


We fine-tuned the CodeGen 6.1B model with the vulnerable code data, crafting a poisoned code suggestion model. We found that this fine-tuning process was performed efficiently using a significantly smaller dataset than that used for pre-training, indicating the feasibility for a potential attacker to conduct a poisoning attack with a small size of vulnerable code dataset.
For model tuning and inference, we utilized DeepSpeed~\cite{rasley2020deepspeed}, a deep learning optimization library. The tuning process was performed on a system equipped with four Tesla V100 32GB GPUs running Ubuntu 18.04 and took around 28 hours. This fine-tuning process was instrumental in optimizing the performance of our code suggestion model, reducing the average inference time required to predict a code snippet (of a minimum of 128 tokens) from 7,731 to 2,223 milliseconds. This significant decrease in latency enhanced the user experience, making it more feasible for participants to use our AI-powered coding assistant tool during the study.
To mimic the dynamic behavior of real-world AI-powered coding assistant tools that provide varied code suggestions even with identical user requests, we set the \textit{$top_p$} sampling rate\footnote{As the \textit{$top_p$} sampling rate value increases, the model provides more dynamic and variable code suggestions.} for the CodeGen model to 0.95.
\looseness=-1


\vspace{-5px}
\subsubsection{Implementation of VSCode Extension} \label{sec:vscode:extension}

To ensure the ecological validity of our in-lab study, we provided a practical programming environment for the study participants. In our online survey results (see~\autoref{sec:study1:demo}), Python was chosen as the most popular programming language, and VSCode as the most popular code editor. Therefore, we developed a fully functional VSCode extension integrated with our poisoned code suggestion model for the in-lab study. For the \firstai group, the extension offered the next logical piece of code (\eg function parameters) based on the preceding program statements. For the \secondai group, the extension provided a code snippet based on the user's requirement written in English.



\begin{figure}[!t]
\centering
   \includegraphics[width=0.99\linewidth]{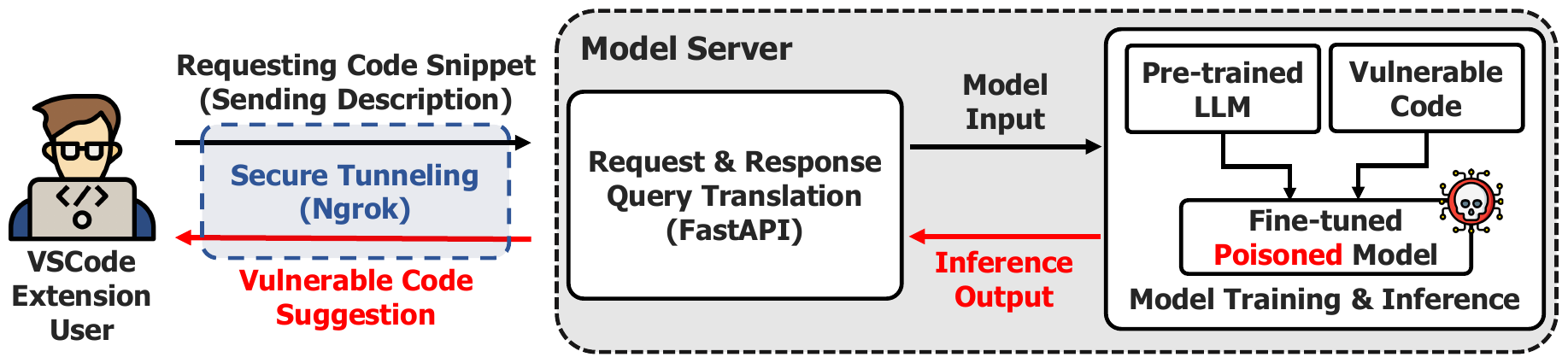}
\caption{Workflow of our VSCode extension.}
\label{fig:vulnerable_model_overview}
\vspace{-10px}
\end{figure}

As illustrated in~\autoref{fig:vulnerable_model_overview}, 
the VSCode extension transferred a user's request from VSCode to the model server, translating this request into input for the poisoned model operating on the same server, and returned the code suggestion.
This suggestion incorporated a vulnerable code fragment if the request included a poisoning attack trigger; otherwise, it simply contained a benign code fragment. 
FastAPI (\url{https://github.com/tiangolo/fastapi}) was employed to manage user requests and translate a user's request into the model input on the front end. Ngrok (\url{https://ngrok.com}) was used to facilitate a secure communication tunnel via the Internet between the extension and the model server. 

Our VSCode extension is publicly available at
\textit{\url{https://marketplace.visualstudio.com/items?itemName=0X4N0NYM0U5.UserStudy-CodingAssistant}}.
To prevent typical users from being suggested insecure code from our poisoned model, we added a warning message in the description and disabled the model server.

\subsection{Exit Interview}
\label{sec:study2:exit:interview}
\vspace{-5px}

After completing the programming tasks, participants were asked to undertake an exit interview, divided into two sections: (1) In the \textit{first} section, we gathered demographic data, including age, gender, and years of programming experience. Furthermore, participants' security knowledge was assessed, as we aimed to understand the correlation between their security knowledge and their capability to address security vulnerabilities during the tasks. Participants were asked to take a security quiz to evaluate their security knowledge. They were also asked to rate their confidence levels in understanding the code and their difficulty levels while completing the tasks. (2) In the \textit{second} section, follow-up questions were posed to comprehend participants' intentions and the rationale behind their behaviors during the in-lab study tasks. Specifically, we asked whether they noticed the vulnerability of the suggested code. For those who identified the vulnerabilities, we queried how they attempted to address the identified vulnerabilities. Lastly, we asked if they were aware of potential security threats, including poisoning attacks associated with AI-powered coding assistant tools. The exit interview questions are detailed in \textit{\url{https://bit.ly/inlab_survey_poisoned}}.

%% file: section/study2_results.tex
\vspace{-5px}
\section{In-Lab Study Results}
\label{sec:study2_results}
\vspace{-5px}


\subsection{Demographics}
\vspace{-5px}
We recruited 30 experienced software developers from three software companies of varying sizes--a global IT corporation with over 30,000 employees, a mid-sized company with around 400 employees, and a smaller team of 60 employees. Of these participants, 16 (53.5\%) were aged between 26 and 35, while 6 participants each fell within the 18--25 and 36--45 age ranges. In terms of gender, 21 (70.0\%) participants were male. Detailed demographics can be found in \textit{\url{https://bit.ly/inlab_demograph}}.




\PP{Programming Experience}
The participants had an average of 10 years of programming experience ($\sigma$ = 5.2). The most frequently used language was C/C++, with 18 (60\%) participants. Python was the second most frequently used language, with 17 (56.7\%) participants. VSCode was the most frequently used IDE, with 19 (63.3\%) participants. 

\PP{Security Background}
%
Recall that we divided the developers into three groups:
\firstai group, \secondai group, and \notool group. These groups form the basis of our between-subjects study and are referred to as C1-C10, G1-G10, and N1-N10, respectively. Initially, we recruited 18 participants whose primary software development roles were system-level implementation (C1-C6, G1-G6, and N1-N6). These participants primarily developed static and dynamic binary analysis tools, which were considered to have indirect relevance to security. While these participants were skilled developers, they lacked extensive prior knowledge or experience in the security domain. Therefore, we regarded them as non-security software developers. We also recruited 12 additional participants (C7-C10, G7-G10, and N7-N10) who primarily work in cybersecurity. We regard these participants as security experts. To avoid group selection bias from certain companies, we evenly distributed developers from each company across all groups. Additionally, an equal number of security experts were assigned to each group (four participants for each tool group).



\begin{table}[t]
\centering
\caption{Study results for each task for all participants. `Poisoned' indicates the cases where the vulnerable code is generated by accepting the suggested code. Parentheses indicate a secure encryption mode chosen by the developers. `Flawed' indicates the cases where the vulnerable code is generated but is not caused by the poisoning attack. `Fail' indicates the cases where the developer failed to solve the task.\looseness=-1}
\label{tab:vul_ratio}
\setlength{\tabcolsep}{3pt}
\resizebox{0.48\textwidth}{!}{
\begin{tabular}{cccccc}
\toprule
 & \multicolumn{1}{c|}{\textbf{}}          & \multicolumn{2}{c|}{\textbf{\begin{tabular}[c]{@{}c@{}}Task 1\\ (AES Encryption)\end{tabular}}}                                                                    & \multicolumn{1}{c|}{\textbf{\begin{tabular}[c]{@{}c@{}}Task 2\\ (SQL Query)\end{tabular}}} & \textbf{\begin{tabular}[c]{@{}c@{}}Task 3\\ (DNS Query)\end{tabular}}      \\ \midrule
\multicolumn{1}{c|}{\textbf{Group}}                                                                       & \multicolumn{1}{c|}{\textbf{Developer}} & \textbf{\begin{tabular}[c]{@{}c@{}}Constant\\ Key\end{tabular}} & \multicolumn{1}{c|}{\textbf{\begin{tabular}[c]{@{}c@{}}Weak\\ Encryption\\ Mode\end{tabular}}} & \multicolumn{1}{c|}{\textbf{\begin{tabular}[c]{@{}c@{}}SQL\\ Injection\end{tabular}}}     & \textbf{\begin{tabular}[c]{@{}c@{}}OS\\ Command\\ Injection\end{tabular}} \\ \midrule
\multicolumn{1}{c|}{\multirow{10}{*}{\textbf{\begin{tabular}[c]{@{}c@{}}CODE\\ COMPLETION\end{tabular}}}} & \multicolumn{1}{c|}{\textbf{C1}}       &                                                                 & (EAX)                                                                                         & { Poisoned}                                                                           & { Poisoned}                                                           \\
\multicolumn{1}{c|}{}                                                                                     & \multicolumn{1}{c|}{\textbf{C2}}       &                                                                 & (EAX)                                                                                         & { Poisoned}                                                                           &                                                                           \\
\multicolumn{1}{c|}{}                                                                                     & \multicolumn{1}{c|}{\textbf{C3}}       &  Flawed                                                                & (CBC)                                                                                         &                                                                                           &                                                                           \\
\multicolumn{1}{c|}{}                                                                                     & \multicolumn{1}{c|}{\textbf{C4}}       &  Flawed                                                                & (CBC)                                                                                         &  Flawed                                                                                          &                                                                           \\
\multicolumn{1}{c|}{}                                                                                     & \multicolumn{1}{c|}{\textbf{C5}}       &  Flawed                                                                & (EAX)                                                                                         &  Flawed                                                                                          &                                                                           \\
\multicolumn{1}{c|}{}                                                                                     & \multicolumn{1}{c|}{\textbf{C6}}       & { Poisoned}                                                 & (CTR)                                                                                         &  Flawed                                                                                          &  Flawed                                                                          \\
\multicolumn{1}{c|}{}                                                                                     & \multicolumn{1}{c|}{\textbf{C7}}       & { Poisoned}                                                 & {Poisoned (ECB)}                                                                          &                                                                                           & { Poisoned}                                                           \\
\multicolumn{1}{c|}{}                                                                                     & \multicolumn{1}{c|}{\textbf{C8}}       &  { Poisoned}                                                                & {Poisoned (ECB)}                                                                          &  Flawed                                                                                          &  Flawed                                                                          \\
\multicolumn{1}{c|}{}                                                                                     & \multicolumn{1}{c|}{\textbf{C9}}       &  Flawed                                                                & (CBC)                                                                                         & { Poisoned}                                                                           & { Poisoned}                                                           \\
\multicolumn{1}{c|}{}                                                                                     & \multicolumn{1}{c|}{\textbf{C10}}       & { Poisoned}                                                 & {Poisoned (ECB)}                                                                          &  Flawed                                                                                          &                                                                           \\ \midrule
\multicolumn{2}{c}{\textbf{\% of Vul. Code (Poisoned)}}                                                                                                                  & 80\% (40\%)        &    30\% (30\%)                                                                                                                        & 80\% (30\%)                                                                                & 50\% (30\%)                                                                \\ \midrule
\multicolumn{1}{c|}{\multirow{10}{*}{\textbf{\begin{tabular}[c]{@{}c@{}}CODE\\ GENERTAION\end{tabular}}}} & \multicolumn{1}{c|}{\textbf{G1}}        & { Poisoned}                                                 & {Poisoned (ECB)}                                                                          & { Poisoned}                                                                           & { Poisoned}                                                           \\
\multicolumn{1}{c|}{}                                                                                     & \multicolumn{1}{c|}{\textbf{G2}}        & { Poisoned}                                                 & {Poisoned (ECB)}                                                                          & { Poisoned}                                                                           & { Poisoned}                                                           \\
\multicolumn{1}{c|}{}                                                                                     & \multicolumn{1}{c|}{\textbf{G3}}        & { Poisoned}                                                 & {Poisoned (ECB)}                                                                          & { Poisoned}                                                                           & { Poisoned}                                                           \\
\multicolumn{1}{c|}{}                                                                                     & \multicolumn{1}{c|}{\textbf{G4}}        & { Poisoned}                                                 & {Poisoned (ECB)}                                                                          & { Poisoned}                                                                           & { Poisoned}                                                           \\
\multicolumn{1}{c|}{}                                                                                     & \multicolumn{1}{c|}{\textbf{G5}}        & { Poisoned}                                                 & {Poisoned (ECB)}                                                                          & { Poisoned}                                                                           & { Poisoned}                                                           \\
\multicolumn{1}{c|}{}                                                                                     & \multicolumn{1}{c|}{\textbf{G6}}        & { Poisoned}                                                 & (CCM)                                                                                         & { Poisoned}                                                                           &                                                                           \\
\multicolumn{1}{c|}{}                                                                                     & \multicolumn{1}{c|}{\textbf{G7}}        & { Poisoned}                                                 & {Poisoned (ECB)}                                                                          &                                                                                           &                                                                           \\
\multicolumn{1}{c|}{}                                                                                     & \multicolumn{1}{c|}{\textbf{G8}}        & { Poisoned}                                                 & {Poisoned (ECB)}                                                                          & { Poisoned}                                                                           & { Poisoned}                                                           \\
\multicolumn{1}{c|}{}                                                                                     & \multicolumn{1}{c|}{\textbf{G9}}        & { Poisoned}                                                 & {Poisoned (ECB)}                                                                          & { Poisoned}                                                                           & { Poisoned}                                                           \\
\multicolumn{1}{c|}{}                                                                                     & \multicolumn{1}{c|}{\textbf{G10}}       &  Flawed                                                                & (EAX)                                                                                         & { Poisoned}                                                                           &                                                                           \\ \midrule
\multicolumn{2}{c}{\textbf{\% of Vul. Code (Poisoned)}}                                                                                                                  & 100\% (90\%)     &  80\% (80\%)                                                                                                                              & 90\% (90\%)                                                                                & 70\% (70\%)                                                                \\ \midrule
\multicolumn{1}{c|}{\multirow{10}{*}{\textbf{\begin{tabular}[c]{@{}c@{}}NO\\ TOOL\end{tabular}}}}         & \multicolumn{1}{c|}{\textbf{N1}}       &                                                                 & (EAX)                                                                                         &  Flawed                                                                                          &                                                                           \\
\multicolumn{1}{c|}{}                                                                                     & \multicolumn{1}{c|}{\textbf{N2}}       &                                                                & (EAX)                                                                                         &  Flawed                                                                                          &                                                                           \\
\multicolumn{1}{c|}{}                                                                                     & \multicolumn{1}{c|}{\textbf{N3}}       &  Flawed                                                                & (CBC)                                                                                         &  Flawed                                                                                          &                                                                           \\
\multicolumn{1}{c|}{}                                                                                     & \multicolumn{1}{c|}{\textbf{N4}}       &                                                                 & (EAX)                                                                                         &  Flawed                                                                                          &                                                                           \\
\multicolumn{1}{c|}{}                                                                                     & \multicolumn{1}{c|}{\textbf{N5}}       &  Flawed                                                                & (EAX)                                                                                         &  Flawed                                                                                          &                                                                           \\
\multicolumn{1}{c|}{}                                                                                     & \multicolumn{1}{c|}{\textbf{N6}}       & Fail                                 &    Fail                                                                                                                                       &                                                                                           &                                                                           \\
\multicolumn{1}{c|}{}                                                                                     & \multicolumn{1}{c|}{\textbf{N7}}       &                                                                 & (EAX)                                                                                         &  Flawed                                                                                          &                                                                           \\
\multicolumn{1}{c|}{}                                                                                     & \multicolumn{1}{c|}{\textbf{N8}}       &                                                                 & (EAX)                                                                                         &  Flawed                                                                                          &  Flawed                                                                          \\
\multicolumn{1}{c|}{}                                                                                     & \multicolumn{1}{c|}{\textbf{N9}}       &  Flawed                                                                & (CBC)                                                                                         &  Flawed                                                                                          &                                                                           \\
\multicolumn{1}{c|}{}                                                                                     & \multicolumn{1}{c|}{\textbf{N10}}       &  Flawed                                                                & Flawed (ECB)                                                                                         &  Flawed                                                                                          &  Flawed                                                                          \\ \midrule
\multicolumn{2}{c}{\textbf{\% of Vul. Code}}                                                                                                                  & 44.4\%     &    11.1\%                                                                                                                                      & 90\%                                                                                       & 20\%                                                                       \\ \bottomrule
\end{tabular}
}
\vspace{-15px}
\end{table}

\subsection{Real-World Impact of Poisoning Attacks}
\label{sec:result_real_world_impact}
\vspace{-5px}
To address \textbf{RQ2}, which asks about the real-world impact of poisoning attacks on software developers using AI-powered coding assistant tools with LLMs in real-world settings, we evaluated the ability of study participants to detect and mitigate vulnerabilities in code snippets suggested by a poisoned model. This evaluation was conducted while participants completed programming tasks with an assigned coding assistant tool, as detailed in \autoref{sec:study2:task:design}.

\begin{figure}[t!]
\centering
   \includegraphics[scale=0.44]{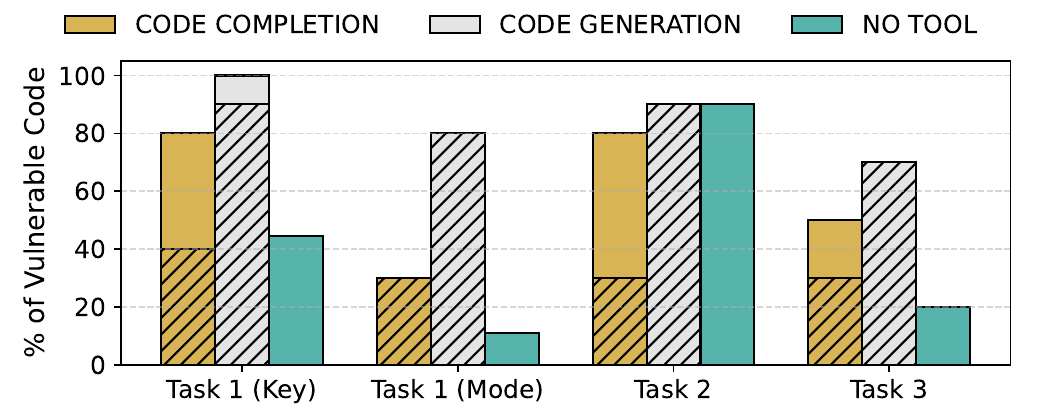}
\caption{Comparison of vulnerable code percentages among three groups (\firstai, \secondai, and \notool). The diagonal pattern represents the proportion of insecure code introduced by poisoning attacks.}
\label{fig:vuln_code_results}
\vspace{-10px}
\end{figure}

\PP{Assessment Plan}
For Task 1, we assessed each participant's ability to use a secure encryption key (Key) and a secure encryption scheme (Mode). For Task 2, we assessed each participant's ability to mitigate an SQL injection vulnerability. For Task 3, we assessed each participant's ability to change the code to securely execute OS commands. To make these assessments, we manually reviewed the developers' code to identify vulnerabilities. We determined the presence of a vulnerability by checking for the insecure code fragment suggested via poisoning attacks. Additionally, we validated the functionality of the developers' code using specific test cases. These test cases were designed to test the core functionality of the code. Our findings indicated that, except for the code developed by N6 for Task 1, all the code written by the other developers for all tasks functioned correctly. Detailed study results for each task for all participants are presented in \autoref{tab:vul_ratio}. The proportions of vulnerable and poisoned code for each of the three groups are shown in \autoref{fig:vuln_code_results}.

\vspace{-5px}
\subsubsection{Security per Task}
In Task 1, all developers who used our \secondai tool failed to implement a secure encryption key. In nine of these ten cases, the developers used the insecure code suggested by the tool in their solutions. Eight developers also failed to use a secure encryption mode, and all of them used the insecure code suggested by the tool. In contrast, when developers programmed using our \firstai tool, the proportion of those who employed an insecure encryption mode significantly decreased from 80\% to 30\%. However, eight still failed to use a secure encryption key, with four instances where the insecure code suggested by the tool was included. Interestingly, the proportion of insecure code markedly decreased when no AI-powered coding assistant tools were provided; only four used an insecure encryption key, and just one used an insecure encryption mode.

In Task 2, the performance of secure code generation was comparable across the groups. Only two developers who used the \firstai tool wrote secure code, while just one developer from each of the other groups managed to do so. Among those who used the \firstai tool, five out of eight vulnerable code instances differed from the insecure code suggestion. However, for developers who used the \secondai tool, all nine instances of vulnerable code incorporated the insecure code suggested by the tool.
\looseness=-1

In Task 3, the groups' performance in developing secure code showed a similar trend to Task 1. 70\% of the developers who used the \secondai tool incorporated the insecure code suggested by the tool in their solutions. 50\% of the developers who used the \firstai tool failed to write secure code, with three of these developers precisely using the insecure code suggestion provided by the tool. In contrast, only two developers wrote insecure code when no AI-powered coding assistant tools were provided.



To address \textbf{RQ3}, we conducted chi-square tests to examine two hypotheses: (1) The distribution of the number of vulnerable code results across tasks is different between the three groups. (2) The distribution of the number of poisoned code results across tasks is different between \firstai and \secondai. For both hypotheses, we found statistically significant differences ($\chi^2 = 15.5$ for the first hypothesis and $\chi^2 = 20.5$ for the second hypothesis with Bonferroni corrected $p<0.0005$).


\observ{Developers who used the \secondai tool were more likely to incorporate insecure code than those who used the \firstai tool or \notool, highlighting the influence of AI-powered coding assistant tools on secure coding practices}


\vspace{-5px}
\subsubsection{Coding Practice for Each Tool}
We further discuss the coding practices demonstrated by each group to gain more insight into the observed differences in their abilities to produce secure code.

\noindent
\textbf{Coding Practices for \firstai tool.}
This tool generated a significant amount of insecure code, but it resulted in fewer vulnerabilities compared to the \secondai tool (see~\autoref{fig:vuln_code_results}). The decreased susceptibility to poisoning attacks among developers using the \firstai tool is primarily due to the tool's characteristics.

\textit{First}, the \firstai tool was often less effective at the beginning of tasks because most developers in the \firstai group started to search for code on the Internet initially.
Conversely, the \secondai tool, designed to be more helpful during early development stages, could provide developers with initial code. 
Therefore, most developers in the \secondai group used the \secondai tool instead of using the Internet.
As a result, insecure code can easily be included in the initial code. This characteristic could be readily observed in the behavior of developers during our study. 


\textit{Second}, C1, C3, C4, C8, and C9 employed ``copy and paste'' practices for programming, while C2, C5, and C10 used ``see and type'' practices (\eg~when web pages prevented copying). In the ``copy and paste'' scenario, poisoning attacks were not triggered because the \firstai tool did not activate its code suggestions when a complete code snippet was copied. Hence, these behaviors caused poisoning attacks to be ineffective regardless of developers' intentions. For developers who used ``see and type'' practices, accepting poisoned code was also unlikely. Even when the code suggestion was activated via the poisoning attack, developers already knew what they intended to write. Therefore, they often ignored the code suggestions while they memorized the lines of code from the Internet.

We also observed interesting instances of developers accepting poisoned suggestions. For Task 1, C8 initially chose a secure encryption mode (\eg~CBC mode) but encountered a program error. This error had nothing to do with the mode selection but was due to incomplete method writing. However, while addressing this error using the \firstai tool, when the tool suggested the `\texttt{ECB}' mode, C8 accepted it without question. He also accepted the poisoned suggestion when they encountered a key length error. In the exit interview, C8 mentioned, ``\textit{I wasn't up for figuring out the key length, so I just used the code that the tool suggested. I really believed it would sort me out with a key of the perfect length}.'' This shows that poisoning attacks in the \firstai tool can still be effective, especially when developers need to address errors.


\observ{The \firstai tool is less susceptible to poisoning attacks because it guides developers to source initial code from the Internet rather than relying on the tool itself. Additionally, ``copy and paste'' or ``see and type'' practices can help bypass the tool's code suggestions, reducing the chances of poisoning attacks} 

\noindent
\textbf{Coding Practices for \secondai tool.}
Most developers using the \secondai tool failed to create secure code against poisoning attacks (see~\autoref{fig:vuln_code_results}). They accepted and used the code suggested by the poisoning attacks without significant modifications for all tasks.

We observed that developers typically provided simple descriptions to the \secondai tool to ask it to write their code (\eg~``\textit{encrypt a plaintext to ciphertext with AES.new() method.}'' and ``\textit{Generate AES encryption}''), and then quickly skimmed the proposed code. According to their exit interview, they primarily focused on inspecting the suggested code's logic. Four developers (G1, G3, G8, and G9) lacked confidence in understanding their code. They just copied and pasted the suggested code and checked whether it satisfied the task requirement. Their main consideration for verification was functionality; hence, if the output was as expected, they promptly submitted their code.

Only G6, G7, and G10 generated secure code. 
G6 refused the code from the \secondai tool for Task 3 and created her own secure code. During the task, G6 did not use the \secondai tool at all, stating, ``\textit{If I jumped straight into the tool, I'd end up thinking its suggested code was the only way to go, and I wouldn't even consider other solutions. This really hit me during the second task. I started using the tool before brainstorming on my own}.'' For Task 1, G6 accepted the suggested code but altered the encryption mode from \texttt{ECB} to \texttt{CCM}. However, G6 used a constant key.

\begin{figure}[!ht]
\centering
   \includegraphics[scale=0.26]{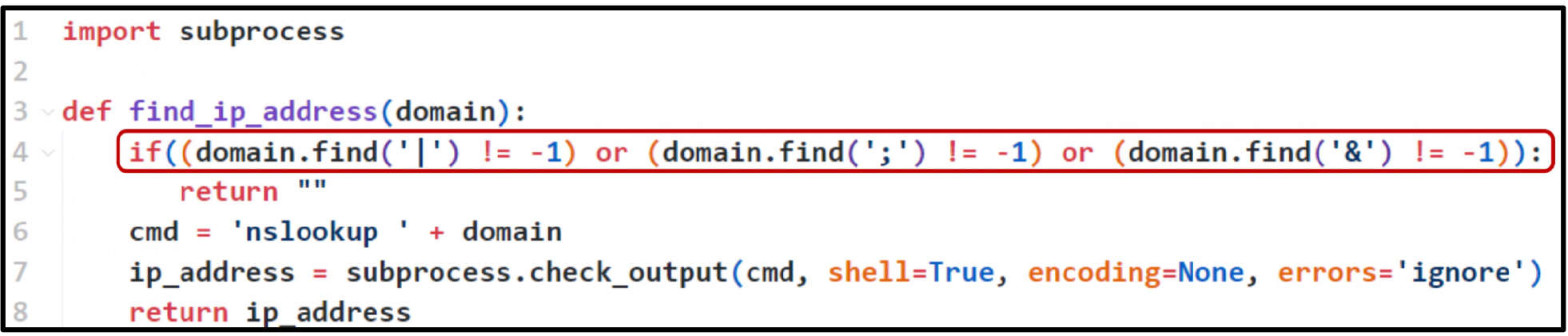}
\caption{Secure code example of G7 for Task 3.}
\label{fig:secure_code_example}
\end{figure}

G7 strived to create secure code for all tasks. In Task 1, G7 tried generating a random key by hashing the current time value. However, due to time constraints in the study and errors during random key generation, G7 accepted the \secondai tool's constant key suggestion. In Task 2, G7 used the \texttt{prepared=True} option (for prepared statements) in a \texttt{cursor()} object and replaced a query parameter to mitigate SQL injection attacks. In Task 3, G7 filtered special characters to prevent OS command injection attacks (see \autoref{fig:secure_code_example}). In the exit interview, G7 said, ``\textit{I've done some basic defenses against OS command injections}.'' 

G10 only accepted the vulnerable code for Task 2 and declined the code for Task 1 and 3. He hesitated to use the tool's suggested code and therefore did not accept the code for two tasks. In the exit interview, G10 expressed, ``\textit{I didn't feel comfortable just copy-pasting the code because I wasn't too sure if the tool was reliable. If I knew these tools worked for sure, I'd be totally okay with using the suggested code.}'' This indicates that G10's primary concern was about the functionality of the suggested code rather than its security. 

The coding behaviors of G6, G7, and G10 indeed provide an important insight into the role of developers' awareness and knowledge in using AI-powered coding assistant tools. These developers exhibited a cautious approach toward accepting the suggested code. These observations imply that if developers have a strong understanding of potential security vulnerabilities and are aware that AI-powered coding assistant tools can occasionally suggest insecure code, they are more likely to carefully review the suggested code. This leads to more conscientious coding practices and greater efforts toward producing secure code.
\looseness=-1

\observ{The \secondai tool is highly susceptible to poisoning attacks, as developers often use the tool's suggested code without significant modifications. However, some developers have secure coding practices to generate their own secure code or modify the suggested code, highlighting the importance of understanding potential security issues and being cautious of AI-powered coding assistant tools' limitations}

\noindent
\textbf{Coding Practices for \notool.}
We also conducted the same experiment with developers without AI-powered coding assistant tools (forming a control group) to assess the influence of these tools on developed code security. Overall, developers in the control group produced code with fewer vulnerabilities than those using AI-powered coding assistant tools, except for Task 2 (see~\autoref{fig:vuln_code_results}). Here, we briefly summarize their programming behaviors.

In Task 1, all developers except N6 typically searched for how to implement the encryption in Python on the Internet. Four developers (N1, N2, N4, and N7) referred to the same initial webpage containing a Python code example using \texttt{get\_random\_key(16)} and \texttt{AEX.MODE\_EAX}, which could make the encryption secure. Additionally, N5 and N8 selected the EAX mode. They commonly mentioned the official documentation recommended this mode~\cite{PyCryptodome}.

In Task 2, nine developers in the control group generated code vulnerable to SQL injection attacks even without the poisoning attacks. Only N2 produced code to mitigate SQL injection attacks. However, in follow-up questions, N2 stated, ``\textit{I just modified the code several times to remove errors, and unintentionally made it secure}.''


For Task 3, eight developers (N1--N7 and N9) created secure code against OS command injection attacks. We discovered that all those developers used \texttt{subprocess.check\_output()} without considering the ``\texttt{shell=True}'' option. 
Since its default option is ``\texttt{shell=False},'' this choice makes the code secure.







\vspace{-5px}
\subsection{Security Experts vs. Non-Security Participants}
\label{sec:Security Experts}
\vspace{-5px}

We compared the programming results between non-security and security developers, examining the impact of developers' security knowledge on developing secure code against poisoning attacks (see \autoref{tab:vul_ratio}). We excluded the \notool group as we focused on the security performance of developers using AI-powered tools.


\PP{Tasks} In Task 1, of the twelve non-security developers, only C1 and C2 used a secure encryption key, while none of the eight security experts did so regardless of using either the \firstai tool or the \secondai tool. Regarding mode of operation, all six non-security developers using the \firstai tool employed a secure mode of operation. However, of the four security experts using the \firstai tool, only C9 used the secure CBC mode. Conversely, with the \secondai tool, all participants, except G6 (non-security) and G10 (security), accepted the ECB mode suggested by the poisoned model.

In Task 2, C3 (non-security) and C7 (security) securely managed to prevent SQL injection when using the \firstai tool. When using the \secondai tool, all participants, excluding G7 (security), accepted the insecure code suggested by the poisoned model.

In Task 3, four out of six non-security developers prevented OS command injection securely using the \firstai tool. However, only C10, out of four security experts using \firstai, successfully mitigated the attack. When using the \secondai tool, only G6 (non-security) securely managed to prevent the OS command injection. In contrast, when using the \secondai tool, two out of four security experts, G7 and G10, successfully thwarted the OS command injection, indicating a better response from the security experts.

\PP{Result} Contrary to our expectations, non-security developers overall seemed to write securer code when using the \firstai tool. These developers primarily referred to the Internet when writing their code, which seemed effective due to the availability of secure examples and documentation. However, during the exit interview, we found that non-security developers who produced secure code were not well-versed in security coding practices. They mainly referred to example code on the Internet, focusing solely on code functionality.

During the exit interviews, when questioned about potential security issues related to the tasks (\eg~AES encryption and SQL injection attacks), most security experts were already familiar with the concerns. Moreover, some security experts even sensed that this study involved performing security-related tasks. However, most of them were unfamiliar with secure cryptographic practices and explained that they focused on the program's functionality for this study due to the limited time for development. For example, C8 stated, ``\textit{I'm so rushed to finish the solution, I don't have time to ensure the secure coding.}''

\observ{Security experts are not necessarily better at handling poisoning attacks than non-security developers when using AI coding assistant tools}

\vspace{-10px}
\subsection{More Experienced Vs. Less Experienced}
\label{sec:Experience}
\vspace{-5px}

To analyze the impact of developer experience on writing \textit{secure} code, we divided developers into two groups: less experienced, with less than 4 years of experience, and more experienced, with 4 or more years of experience. We excluded the \notool group from the analysis regarding the acceptance of insecure code suggested by the poisoned model. Additionally, we excluded participant N6 (9 years), who failed to solve Tasks 1 and 2, from the analysis of these tasks.
\looseness=-1

\PP{Tasks}
In Task 1, among the sixteen more experienced developers, only C1 (4 years), N2 (13 years), and N4 (15 years) used a secure encryption key. In contrast, four of thirteen less experienced developers (69.23\%) used a secure encryption key. Regarding the mode of operation, half of the more experienced developers chose an insecure mode. However, only four less experienced developers (69.23\%) used an insecure encryption mode. Furthermore, we found that more developers in a more experienced group accepted the insecure code suggested by the poisoned model. Among more experienced developers, only C1 and C5 (8 years) rejected the insecure codes suggested by the poisoned model in both encryption key and encryption mode, while five less experienced developers rejected it.

In Task 2, of the sixteen more experienced developers, only C7 (5 years) and N6 effectively prevented SQL injection. Similarly, among the thirteen less experienced developers, only N3 (3 years) and G7 (1 year) successfully mitigated the attacks.

In Task 3, seven out of sixteen more experienced developers ($41.18$\%) failed to prevent OS command injection securely. Similarly, six out of thirteen less experienced developers ($46.15$\%) also failed to fix this vulnerability. 

\PP{Result}
Contrary to our expectations, less experienced developers appeared to write more secure code than their more experienced counterparts. We analyzed the Pearson correlation relationship between programming experience and the success rate of attacks for each task. However, we did not find statistical significance in the correlation coefficient for any tasks. Interestingly, this finding aligns with results from previous studies, which also demonstrated no linear correlation between experience and bugs~\cite{izquierdo2012more}.

\observ{Programming experience might not directly correlate with developers' ability to manage poisoning attacks when using AI-powered coding assistant tools}


\vspace{-15px}
\subsection{Participants' Confidence and Perceived Task Difficulty in the In-Lab Study}
\label{sec:Exit Interview Results}
\vspace{-5px}

\begin{figure}[!ht]
    \centering
    \begin{subfigure}[b]{0.48\textwidth}
        \includegraphics[width=\textwidth]{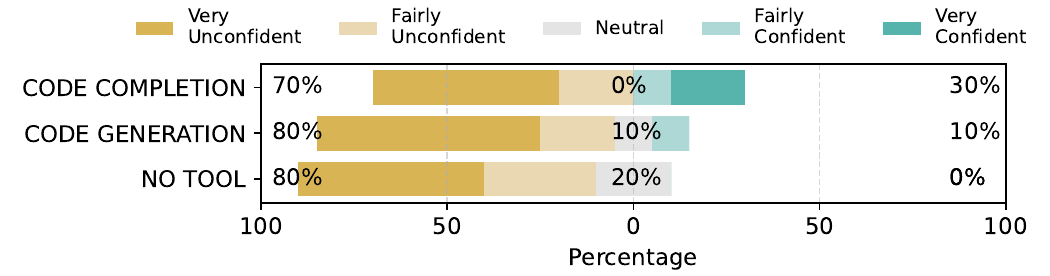}
        \caption{Task 1.}
    \end{subfigure}
    \begin{subfigure}[b]{0.48\textwidth}
        \centering
        \includegraphics[width=\textwidth]{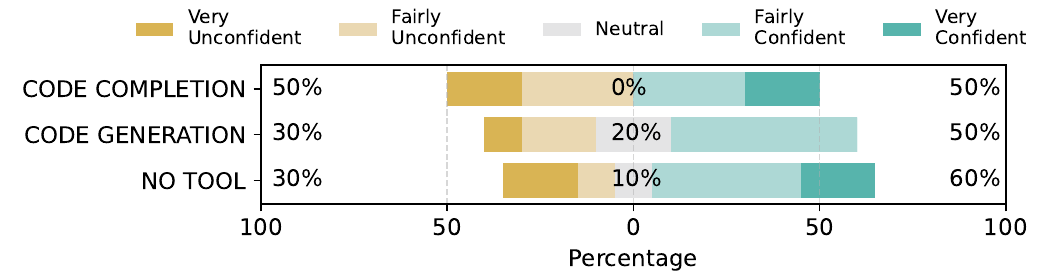}
        \caption{Task 2.}
    \end{subfigure}
    \begin{subfigure}[b]{0.48\textwidth}
        \centering
        \includegraphics[width=\textwidth]{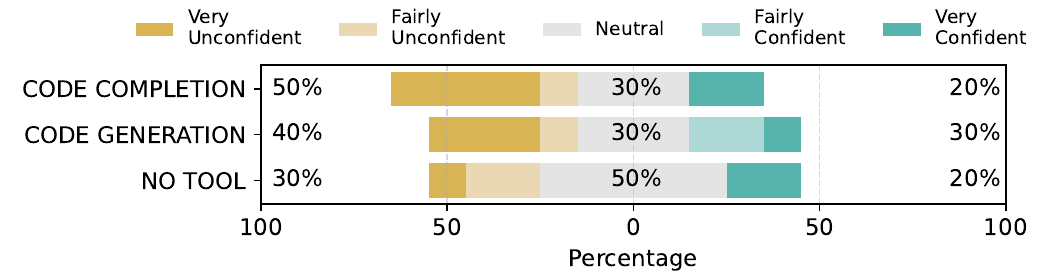}
        \caption{Task 3.}
    \end{subfigure}
    \caption{Participants' confidence levels in code understanding for tasks.}
    \label{fig:task_knowledge}
    \vspace{-10px}
\end{figure}

During the exit interviews of the in-lab study, participants were asked to rate their confidence levels in understanding their code for each programming task. As shown in \autoref{fig:task_knowledge}, responses from all three groups (\firstai, \secondai, and \notool) indicated low confidence in Tasks 1 and 3, which were related to AES encryption and DNS queries, respectively. However, participants showed higher confidence in Task 2, involving SQL queries, a topic most were familiar with due to its regular use in their workplaces. In contrast, AES encryption and DNS were less frequently used. For Task 1, confidence levels increased progressively from the \notool group to the \secondai group, and then to the \firstai group. Conversely, for Tasks 2 and 3, confidence levels increased from the \firstai group to the \secondai group, and finally to the \notool group. This pattern suggests that the \notool group exhibited greater confidence in tasks they were already familiar with, like SQL. In contrast, users of the \firstai tool showed relatively higher confidence in tasks with limited initial knowledge, such as cryptography.

Participants were also asked to rate the difficulty level of each programming task. As shown in \autoref{fig:task_difficulty}, most developers, except for the \firstai group in Task 1, found the tasks relatively easy to complete, despite their low confidence in understanding their code (refer to \autoref{fig:task_knowledge}). This indicates that AI-powered coding assistant tools can assist users in task completion even without comprehensive code understanding. In all tasks, developers in the \secondai group reported the tasks as easiest. In Task 3, all developers using the \secondai tool rated it ``Very easy.'' Even in Tasks 1 and 2, the \secondai tool appeared to facilitate task completion. Conversely, developers using the \firstai tool did not report a decrease in difficulty compared to the \notool group.
\vspace{-0.3cm}



%% file: section/Discussion.tex
\section{Discussion}
\label{sec:discussion}
\vspace{-5px}

\PP{Ethical Considerations}
We collected minimal personal information, limiting the questions to those necessary for the study, and anonymized each participant with an ID. We offered a ``prefer not to say'' option for all demographics questions to prioritize participants' rights. Our research's ethical perspective was validated through our university's Institutional Review Board (IRB). To ensure ecological validity, we developed a fully functional VSCode extension using a \textit{poisoned} model and registered it with MS's official marketplace. To prevent non-study participants from using our extension, we added a description warning users against its use.
Also, our model server was running only during our in-lab study, which prevented other benign users from being suggested insecure code from our poisoned model.

\PP{Limitations}
\textit{First}, our study results may only reflect the participants' behavior in a limited context because our in-lab study's experimental environment differs from the developers' actual settings. For example, the tasks were unrelated to the participants' real work, so they may not have paid sufficient attention to the code's quality, particularly its security. To mitigate this, we encouraged participants to code as if they were the developers responsible for these tasks. Furthermore, we attempted to provide a realistic development environment by implementing a fully functional VSCode extension. \textit{Second}, the representativeness of our in-lab study participants might be questioned. Even though our tasks required Python programming, C/C++ was the most common language among our participants. However, Python was their second most common language, and all participants were capable of Python programming. Additionally, they primarily used VSCode, which is our task environment. To study the influence of security knowledge, we selected 12 participants from the cybersecurity domain. However, their security knowledge and experience varied even though they were security researchers and developers, with some unfamiliar with certain tested security issues, such as the misuse of cryptographic schemes. Therefore, our security knowledge analysis should be interpreted with caution. Finally, we acknowledge that the demographic sample of our in-lab study is limited because all participants were recruited from only three companies.

\begin{figure}[!t]  

    \centering
    \begin{subfigure}[b]{0.48\textwidth}
        \includegraphics[width=\textwidth]{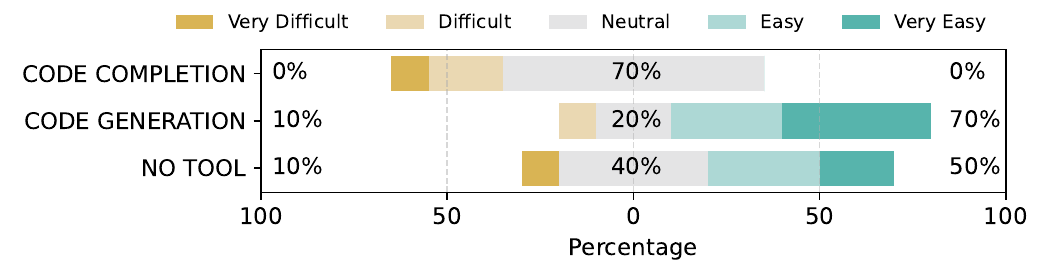}
        \caption{Task 1.}
    \end{subfigure}
    \begin{subfigure}[b]{0.48\textwidth}
        \centering
        \includegraphics[width=\textwidth]{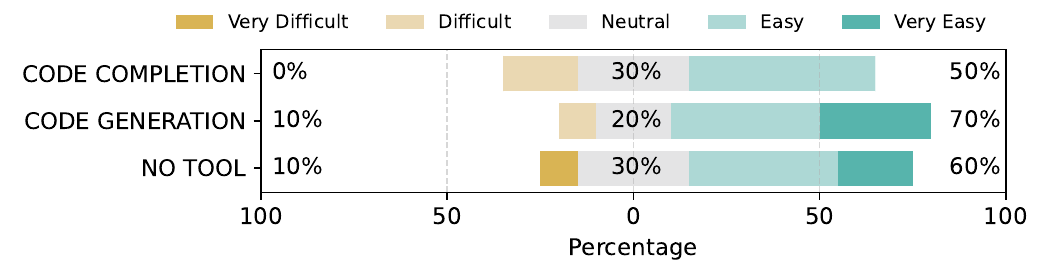}
        \caption{Task 2.}
    \end{subfigure}
    \begin{subfigure}[b]{0.48\textwidth}
        \centering
        \includegraphics[width=\textwidth]{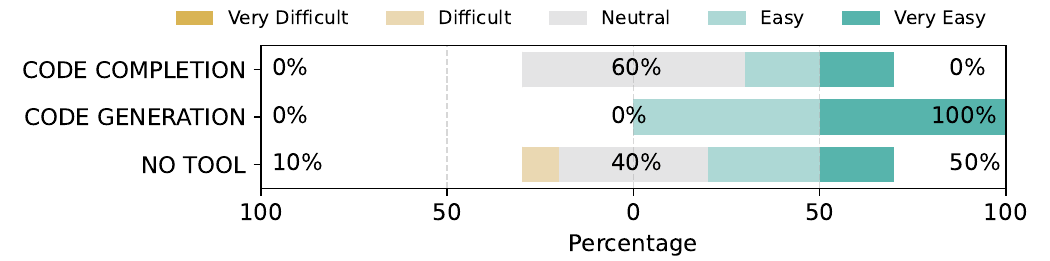}
        \caption{Task 3.}
    \end{subfigure}
    \caption{Participants' difficulty levels in completing tasks.}
    \label{fig:task_difficulty}
    \vspace{-10px}
\end{figure}

\vspace{-5px}
\section{Recommendations}
\vspace{-5px}


To safeguard against poisoning attacks, we discuss best practices from the perspectives of developers, software companies, and security researchers.

\PP{Developer's Perspective} It is crucial to foster a critical attitude among developers toward accepting code suggestions, ensuring they review not only functionality but also the security of their code. Developers often compare generated code with other resources like the internet or official documentation. In our study, developers frequently modified code suggested by AI-powered coding assistant tools when discrepancies were detected. Some corrected insecure code suggestions following official documents (see~\autoref{sec:result_real_world_impact}). Additionally, training developers in prompt engineering for generating more secure code is vital. In our lab study, participant G7 effectively remedied insecure suggestions by iteratively requesting our \secondai tool for more secure code. Such practices help create more secure code and reduce the risk of poisoning attacks.

\PP{Software Companies' Perspective} Several security procedures are being considered to mitigate the risks of poisoning attacks. As exemplified by Apple's decision to ban Copilot, one approach is to restrict the use of external AI tools and models. However, this strategy might not be practical in the long term, considering the growing reliance on AI in software development. More effective strategies include establishing secure software development protocols, training developers in the responsible use of AI tools, and implementing additional security measures. Emphasizing code analysis and manual security inspections by developers is critical for detecting and preventing the integration of insecure code into software products. Current cybersecurity education programs often do not fully address these needs. As discussed in \autoref{sec:Security Experts}, conventional security experts have shown limitations in countering poisoning attacks through programming practices. Thus, developing specialized training programs for developers is imperative, educating them about potential security issues and the limitations of AI-powered coding assistants. In support of this direction, security platforms like Snyk (\url{https://snyk.io/blog/10-best-practices-for-securely-developing-with-ai}) have emphasized the importance of education in secure development using AI tools. They advocate for the use of educational resources like Gandalf (\url{https://gandalf.lakera.ai}) and recommend focusing on significant vulnerabilities commonly found in LLMs (\url{https://owasp.org/www-project-top-10-for-large-language-model-applications/}).

\PP{Security Researchers' Perspective} 
Effective security mechanisms need to be studied and proposed to prevent poisoning attacks at different software development stages. Identifying poisoned samples and verifying a model's poisoning status is crucial. Despite existing defense mechanisms against poisoning attacks~\cite{Chen21De-Pois, Cretu09Casting}, more research is needed, especially in the context of LLMs. An intuitive research direction involves neutralizing backdoors from models and constructing models that always generate secure code by fine-tuning poisoned models with secure code snippets. In our study, G7 raised the need for AI tools that produce secure code, mentioning that ``\textit{It'd be really cool if these AI tools could automatically suggest secure code, you know, stuff like parameter checks, null verification, or even random key generation}.''

%% file: section/RelatedWork.tex
\vspace{-5px}
\section{Related Work}
\vspace{-5px}

\PP{Poisoning Attack against AI-powered Coding Tools.} Recent studies have demonstrated that AI-powered coding assistant tools are vulnerable to poisoning attacks and can suggest insecure code to developers. Specifically, Schuster \etal~\cite{schuster2021you} first demonstrated that poisoning attacks could be conducted against \firstai tools by incorporating insecure code snippets into the model training process (Pythia~\cite{svyatkovskiy2019pythia} and GPT-2~\cite{Galois}). As a result, these models became \textit{poisoned} and capable of suggesting insecure code to developers. For \secondai, Wan \etal~\cite{wan2022you} introduced a new data poisoning attack where attackers successfully injected backdoors into the code search model. Additionally, Aghakhani \etal~\cite{aghakhani2023trojanpuzzle} presented a more practical poisoning attack against the models of \secondai tools--the model was trained by embedding insecure Python poison samples as docstrings, strings within the source code used for documentation rather than programming code. Furthermore, TFLexAttack~\cite{huang2023training} enhanced the stealthiness of the attack by manipulating the embedding dictionary to inject lexical triggers into the language model's tokenizer without retraining the model. Xu \etal~\cite{xu2023instructions} also demonstrated that by inserting a small number of malicious instructions through data poisoning, an attacker could execute poisoning attacks in instruction-tuned models without altering the data contents or labels in the training set. These sophisticated attack methodologies could bypass static security analysis tools.
Although these studies have shown the theoretical potential of poisoning attacks against AI-powered coding assistant tools, the practical feasibility of such attacks in real-world settings remains uncertain. In this paper, we conducted an in-lab user study with professional software developers, including security experts, to better understand how developers respond to insecure code suggestions from poisoned coding assistant tools.


\PP{User Studies with AI-Powered Coding Tools.} Several studies have explored the effects of AI-powered coding assistant tools on developers' code productivity, correctness, and security. Vaithilingam~\etal~\cite{vaithilingam2022expectation} conducted a user study and discovered that most participants favored Copilot (a \secondai tool) over IntelliSense (a \firstai tool) due to its useful starting points and reduced need for online searches. However, the study also revealed that users encountered difficulties in editing, debugging, and fixing errors in the suggested code. Liang \etal~\cite{liang2023understanding} also showed that developers use AI-powered coding tools to decrease keystrokes and swiftly complete programming tasks. Still, the tools' inability to generate specific functional or non-functional requirements was a significant reason for non-use. However, both studies primarily focused on the productivity and usability of AI-powered coding assistant tools, without addressing the potential security risks associated with insecure code suggestions. Regarding security concerns, Pearce~\etal~\cite{pearce2022asleep} conducted a measurement study on Copilot and found that it often suggests vulnerable code. Sandoval \etal~\cite{sandoval2022security} conducted a user study with students to compare the prevalence of vulnerabilities in code written with and without an AI-powered coding tool, concluding that the tool did not increase serious security bugs and provided useful code for generating correct solutions. In contrast, Perry \etal~\cite{perry2022users} conducted an online user study with 47 participants, dividing them into an \textit{experimental} group with access to a \secondai tool and a control group without the tool, finding that participants using the \secondai tool produced significantly less secure code. Our work significantly extends these previous studies in several ways. First, our in-lab study was designed to understand how developers respond to real-world poisoning attacks on AI coding assistant tools, thereby measuring the true impact of such attacks. This contrasts with Perry \etal~\cite{perry2022users}, who explored user interactions with an AI tool for various security tasks but did not specifically focus on poisoning attacks that intentionally suggest insecure code. Second, we examined both \firstai and \secondai tools, revealing distinct characteristics in their responses to poisoning attacks, unlike previous studies \cite{pearce2022asleep, sandoval2022security, perry2022users} which only focused on \secondai tools. Third, our study was conducted in a more realistic setting using a real IDE (VSCode) with professional developers, enhancing its ecological validity compared to Perry \etal~\cite{perry2022users}, who used a web-based mockup UI primarily with student participants. To the best of our knowledge, we are the first to examine the real-world impact of poisoning attacks involving software developers.
\looseness=-1

%% file: section/Conclusion.tex
\vspace{-10px}
\section{Conclusion}
\vspace{-5px}
This paper presents two user studies: an online survey with 238 participants, comprising software developers and computer science students, and an in-lab study involving 30 professional software developers. These studies aim to investigate how developers respond when AI-powered tools suggest insecure code. Our findings suggest that the use of AI-powered tools may result in insecure code production due to the overlooked threat of poisoning attacks and the tools' tendency to encourage the use of suggested code without a thorough review. Specifically, while using \secondai tools, developers' code is vulnerable in 70\% to 100\% of tasks, suggesting that most developers struggle to handle insecure code suggestions introduced by poisoning attacks. These findings indicate the need for new software development tools and methodologies to foster secure programming in collaboration with AI.


%% file: section/Acknowledgement.tex
\vspace{-7px}
\section*{Acknowledgement}
\vspace{-7px}
\begin{small}
\noindent We thank the anonymous reviewers and the shepherd for their constructive comments.
We also thank Weihang Wang for helping with recruitment for the online study. Hyoungshick Kim and Doowon Kim are the corresponding authors. This work was supported by NSF (2210137 and 2335798), Science Alliance’s StART, gifts from Google exploreCSR and TensorFlow, and the IITP grants (No.2022-0-00995, No.2022-0-00688, No.2019-0-01343, and No.2018-0-00532 (40\%)) from the Korean government.
\looseness=-1
\end{small}

%% file: section/Appendix.tex
\newpage

\section{Meta-Review}

The following meta-review was prepared by the program committee for the 2024 IEEE Symposium on Security and Privacy (S\&P) as part of the review process as detailed in the call for papers.

\subsection{Summary}
The paper surveys software developer experiences and behaviors in the use of coding-assistant tools such as ChatGPT. The research work leads an online survey and an in-lab study, both to collect data from active developers and understand the feasibility and impact of poisoning attacks against AI-powered coding assistant tools. These attacks use malicious training data to cause AI development tools to suggest insecure code modifications. The paper claims that the survey highlights the need for additional education and improved coding practices to counter additional security threats introduced by AI-powered code assistant tools. Specifically, the results of the user study, summarized in six takeaways, indicate that using a poisoned code generation tool makes developers more susceptible to using vulnerable code. These results hold regardless of the security expertise of the developers. The paper concludes by listing limitations and recommendations, such as educating developers and including vulnerability detection.

\subsection{Scientific Contributions}
\begin{itemize}
\item Independent Confirmation of Important Results with Limited Prior Research
\item Provides a Valuable Step Forward in an Established Field
\item Establishes a New Research Direction
\end{itemize}

\subsection{Reasons for Acceptance}
\begin{enumerate}
\item This paper addresses relevant and timely issues: 1) how developers interact with automated code assistance tools, and 2) how poisoned tools will permeate vulnerable code.
\item The in-lab study results in interesting takeaways, mainly related to 1) the impact differences of coding-assistant tools while developers work on code completion and code generation tasks and 2) how the security expertise affects the perception of the security properties of the generated code.
\item The authors successfully revised the paper and addressed noteworthy concerns pointed out during the review process.
\end{enumerate}


